\documentclass[twocolumn]{aastex61}

\begin{document}

\title{Persistence of precursor waves in two-dimensional relativistic
shocks}

\author{Masanori Iwamoto} 
\affiliation{Department of Earth and Planetary Science, University of Tokyo,
7-3-1 Hongo, Bunkyo-ku, Tokyo 113-0033, Japan}

\author{Takanobu Amano} 
\affiliation{Department of Earth and Planetary Science, University of Tokyo,
7-3-1 Hongo, Bunkyo-ku, Tokyo 113-0033, Japan}

\author{Masahiro Hoshino} 
\affiliation{Department of Earth and Planetary Science, University of Tokyo,
7-3-1 Hongo, Bunkyo-ku, Tokyo 113-0033, Japan}

\author{Yosuke Matsumoto} 
\affiliation{Department of Physics, Chiba University, 1-33 Yayoi, Inage-ku,
Chiba, Chiba 263-8522, Japan}

\correspondingauthor{Masanori Iwamoto}
\email{iwamoto@eps.s.u-tokyo.ac.jp}

\begin{abstract}

 We investigated the efficiency of coherent upstream large-amplitude
 electromagnetic wave emission via synchrotron maser instability at
 relativistic magnetized shocks by using two-dimensional
 particle-in-cell simulations. We considered the purely perpendicular
 shock in an electron-positron plasma. The coherent wave emission
 efficiency was measured as a function of the magnetization parameter
 $\sigma$, which is defined by the ratio of the Poynting flux to the
 kinetic energy flux. The wave amplitude was systematically smaller than
 that observed in one-dimensional simulations. However, it continued to
 persist, even at a considerably low magnetization rate, where the
 Weibel instability dominated the shock transition. The emitted
 electromagnetic waves were sufficiently strong to disturb the upstream
 medium, and transverse filamentary density structures of substantial
 amplitude were produced. Based on this result, we discuss the
 possibility of the wakefield acceleration model for the production of
 non-thermal electrons in a relativistic magnetized ion-electron shock.
 
\end{abstract}

 \section{Introduction}\label{sec:intro}

 The origin of cosmic rays (CRs) has been a long-standing problem in
 astrophysics. CRs with energies up to $10^{15.5}$ eV are commonly
 believed to be generated by diffusive shock acceleration (DSA)
 \citep{Bell1978,Blandford1978,Drury1983,Blandford1987} at supernova remnant
 shocks in our galaxy. Observations of broadband non-thermal emission
 from young supernova remnants support this paradigm
 \citep[e.g.,][]{Koyama1995, Bamba2006, Helder2009}. In contrast, the
 origin of high-energy CR population ($> 10^{15.5}$ eV), presumably of
 extragalactic origin, is not well understood. Active galactic nuclei
 (AGN) and gamma-ray bursts (GRBs) are among the possible sources of
 such ultra-high-energy CRs (UHECRs) \citep[e.g.,][]{Hillas1984,
 Milgrom1995, Vietri1995, Waxman1995}. Observations show that AGN
 produce high-energy photons via synchrotron and inverse Compton
 emission \citep[e.g.,][]{Abdo2010b, Abdo2010a, Nolan2012}, which
 clearly shows the presence of ultrarelativistic electrons. AGN are
 typically associated with relativistic jets with Lorentz factors of a
 few tens \citep[e.g.,][]{Lister2016}. The energetic electrons are often
 assumed to be produced at shockwaves associated with the jet
 \citep[e.g.,][]{Marscher2008}. The intense radiation from GRBs is also
 thought to originate from a relativistic outflow whose Lorentz factor
 may exceed several hundreds \citep[e.g.,][]{Peer2007,
 Ackermann2014}. The standard model for GRBs assumes the presence of 
 shockwaves wherein charged particles are accelerated and produce
 photons through synchrotron radiation \citep[see, e.g.,][]{Piran2005}. 

 DSA has also been one of the most commonly proposed mechanisms for
 producing UHECRs in these sources. However, it is well known that
 DSA becomes less efficient at a relativistic shock propagating in a
 magnetized plasma \citep{Begelman1990}. The Lorentz transformation of
 the magnetic field from the upstream rest frame into the shock rest
 frame boosts only the transverse component by the Lorentz factor of the
 shock. Therefore, a highly-relativistic shock is almost always
 superluminal, indicating that particles moving along the magnetic field
 line cannot diffuse back into the upstream. In such a shock, particles
 must diffuse {\it across} the magnetic field to be accelerated by the DSA
 mechanism. Since perpendicular diffusion is typically much slower
 than parallel diffusion, this may not be a plausible solution for
 particle acceleration in highly-relativistic shocks \citep[see,
 e.g.,][]{Takamoto2015}. 

 \cite{Chen2002} proposed an alternative model for UHECR acceleration at
 GRBs. The model invokes  
 a large-amplitude Alfv\'en wave in a relativistic plasma that generates an
 electrostatic wave behind it. A particle may be accelerated by this
 electric field to high energies, possibly to the UHECR-energy
 range. This is essentially an application of wakefield acceleration
 (WFA), as discussed in the context of laser-plasma interactions in the
 laboratory \citep{Tajima1979}. WFA in laboratory plasmas occurs when an
 intense laser pulse (or transverse electromagnetic wave) propagates
 in a plasma. A Langmuir wave is excited via Raman scattering, in
 which the ponderomotive force exerted by the laser pulse expels
 electrons from the region of high laser intensity, whereas ions are
 nearly unaffected. Consequently, a large-amplitude charge density
 fluctuation is generated, which is associated with a longitudinal
 electric field called a wakefield.  \added{It is easy to confirm that}
 the phase velocity of the excited Langmuir wave is nearly equal to the
 group velocity of the laser pulse, and is close to \added{but less
 than} the speed of light \added{\citep[see, e.g.,][]{Hoshino2008}}. Therefore,
 the wakefield is able to accelerate particles to highly-relativistic
 energies via the Landau resonance. 

 WFA may also be applicable to relativistic magnetized
 shocks in nature. It is known that large-amplitude electromagnetic
 waves are excited at relativistic shock fronts by synchrotron maser
 instability (SMI) driven by particles reflected off the
 shock-compressed magnetic field \citep{Hoshino1991}. The instability
 results from the resonance between the relativistic particle cyclotron
 motion and an electromagnetic wave of the extraordinary mode (X-mode). The
 fluctuating magnetic field of the X-mode wave is parallel to the
 ambient magnetic field and perpendicular to the wavenumber vector. In a
 pair plasma, the fluctuating electric field is perpendicular to both
 the fluctuating magnetic field and the wavenumber vector and thus the
 X-mode wave is linearly polarized. The electromagnetic waves may be
 emitted toward both the upstream and downstream directions; however, only
 the waves with upstream-directed group velocities greater than
 the shock propagating speed can escape upstream. Consequently,
 the appearance of such high-frequency precursor waves ahead of the
 shock is a common feature of relativistic magnetized shocks
 reproduced by one-dimensional (1D) particle-in-cell (PIC) simulations
 \citep[e.g.,][]{Langdon1988, Gallant1992, Hoshino1992, Amato2006}. The
 precursors emitted in 1D PIC simulations of relativistic magnetized
 shocks in electron-ion plasmas were sufficiently strong to generate a
 wakefield of substantial amplitude wherein electrons were accelerated
 \citep{Lyubarsky2006, Hoshino2008}. \cite{Hoshino2008} demonstrated the
 generation of non-thermal electrons by the WFA and estimated the
 maximum attainable energy $\varepsilon_{max}$ as
 $\varepsilon_{max}/\gamma_1m_ec^2 \sim L_{sys}/(c/\omega_{pe})$, where
 $\gamma_1$ is the upstream bulk Lorentz 
 factor and $\omega_{pe}$ is the proper electron plasma
 frequency. According to this estimate, the maximum energy is
 proportional to the system size $L_{sys}$ of an astronomical object.  
 \cite{Hoshino2008} found that the maximum energy can exceed the
 theoretical limit of conventional shock acceleration models such as DSA
 and that the WFA plays an important role in relativistic magnetized shocks,
 especially at a low magnetization rate.
 
 WFA in the context of relativistic shocks has so far
 been discussed solely through 1D simulations. It is not well known
 whether the same mechanism can operate in more realistic 
 multidimensional systems. In general, inhomogeneity (or waves) may
 appear in the transverse direction of the shock and the waves emitted
 from different positions at the shock may not necessarily be coherent
 in phase. Because the precursor waves are a superposition of such
 waves, the efficiency of the WFA may deteriorate in an incoherent
 precursor. Another possible problem is the competition between the SMI
 and Weibel instability (WI) \citep{Weibel1959, Fried1959}. The WI is driven
 unstable by effective temperature anisotropy generated by the reflected
 particles in the shock-transition region
 \citep[e.g.,][]{Kato2007,Chang2008}. The mode is unstable for wavenumbers
 perpendicular to the shock normal, thereby appearing only in
 multidimensional simulations. The maximum growth rate of the WI
 including relativistic effects scales as the proper electron plasma frequency 
 $\omega_{pe}$ for sufficiently strong anisotropy \citep[see, e.g.,
 ][]{Yoon1987, Yang1993, Achterberg2007, Schaefer2008}. In contrast, the
 growth rate of the SMI is on the order of the relativistic electron
 cyclotron frequency $\omega_{ce}$ \citep{Hoshino1991}. As both
 instabilities are excited from the same free-energy source, the
 precursor wave emission efficiency may be affected or even
 completely shut off in a low-magnetization regime ($\omega_{ce} \ll
 \omega_{pe}$), where the WI
 grows more quickly than the SMI. Previous PIC simulation studies in
 multiple dimensions indeed showed that the shock transition is dominated
 by the WI at low magnetization $\sigma \lesssim 10^{-2}$ \citep[e.g.,
 ][]{Spitkovsky2005, Kato2010, Sironi2013}, where 
 $\sigma$ is the ratio of the Poynting flux to the kinetic energy flux.

 In fact, some earlier studies of two-dimensional (2D)
 magnetized shocks gave negative results for the WFA. \cite{Sironi2009}
 reported that the precursor waves were seen only in the initial phase
 and had little effect in their long-term calculations. Later,
 \cite{Sironi2011} found that although a wakefield was induced in their 2D
 simulations of magnetized shocks in an ion-electron plasma,
 the amplitude was not sufficiently strong to produce non-thermal
 particles, which is in clear contrast to the 1D results reported by
 \cite{Hoshino2008}. However, according to our numerical convergence study (see 
 Appendix \ref{sec:conv}), the numerical resolutions used in
 the earlier studies were insufficient to capture the precursor
 waves. In this study, we quantify the 
 efficiency of precursor wave emission 
 using high-resolution 2D PIC simulations to investigate 
 the applicability of the WFA model to astrophysical
 relativistic magnetized shocks. For the purpose of estimating the
 precursor wave emission efficiency, we consider only
 shocks in electron-positron plasmas. We note that, although the
 WFA needs a finite inertial difference between the
 positive and negative charges, the emission mechanism itself is
 identical between the pair and ion-electron plasmas. Therefore, the
 precursor wave emission efficiency measured in a pair plasma shock will
 also give a good estimate for an ion-electron plasma. 

 This work is organized as follows. First, we describe our
 simulation setup in 
 $\S$\ref{sec:setup}. $\S$\ref{sec:shock} describes global structures
 of relativistic magnetized shocks in the presence of
 large-amplitude electromagnetic precursor waves.  
 In $\S$\ref{sec:precursor}, the properties of precursor
 waves are comprehensively analyzed. In $\S$\ref{sec:discussion}, we discuss the
 applicability of the WFA model to relativistic magnetized shocks in an
 ion-electron plasma. Finally, $\S$\ref{sec:summary} summarizes this study.

 \section{Simulation setup}\label{sec:setup}
 
\begin{figure}[htb!]
  \plotone{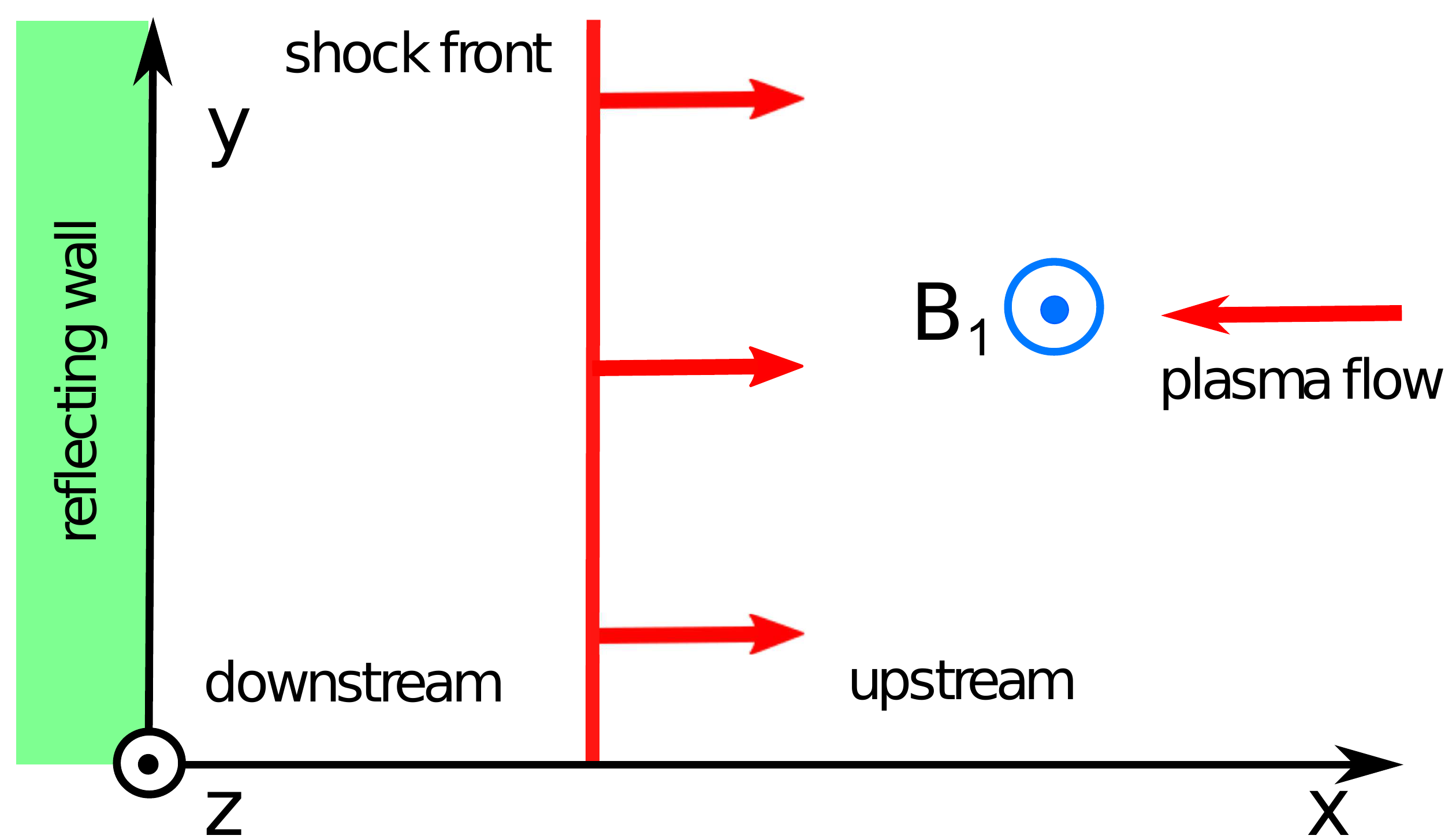}
  \caption{Coordinate system and geometry of the simulation.}
  \label{configuration}
 \end{figure}

 Our simulations of relativistic shocks in magnetized pair plasmas were
 performed in 2D systems in the $x$-$y$ plane using a fully relativistic
 electromagnetic PIC code \citep{Matsumoto2013, Matsumoto2015}. The 
 basic configuration of our simulations is illustrated in Figure
 \ref{configuration}. 
 A cold pair plasma flow (with zero thermal spread) is continuously
 injected from the right-hand boundary of the simulation domain toward
 the $-x$ direction at a bulk Lorentz factor of $\gamma_1=40$.
 The particles are specularly reflected at the left-hand boundary, and the
 conducting-wall boundary condition is applied for the electromagnetic
 field. The boundary condition in the $y$ direction is periodic for both
 the field and the particles. The particles reflected off the wall interact 
 with the incoming plasma, and a shockwave propagating in the
 $+x$ direction is formed. Consequently, the simulation frame corresponds
 to the downstream rest frame.

 In this study, we solely focus on the purely perpendicular shock. In 2D
 simulations, there is a degree of freedom in the choice of direction
 of the ambient magnetic field. In general, due to the neglect of
 the third coordinate ($z$ in our coordinate system), the rotation of
 the ambient magnetic field around the shock normal may change the shock
 dissipation physics. Nevertheless, we only present simulation results
 with the ambient magnetic field pointing in the out-of-plane direction
 ($B_z$ in our coordinate system) in this study for simplifying our
 discussion. Note that our conclusion is not limited to this particular
 configuration. The results with an in-plane magnetic field 
 configuration ($B_y$) will be published elsewhere. 

 The length and time were measured in units of the electron inertial
 length $c/\omega_{pe}$ and the inverse proper electron plasma frequency
 $\omega_{pe}^{-1}$, respectively. The proper electron plasma frequency
 is defined as follows: 
 \begin{equation}
  \label{eq:wpe}
  \omega_{pe}=\sqrt{\frac{4 \pi N_1 e^2}{\gamma_1 m_e}},
 \end{equation}
 where $N_1$ is the upstream electron number density measured in the
 simulation frame. The number of particles/cell is $N_1 \Delta x^2 = 64$
 for both electrons and positrons upstream, where $ \Delta x =
 \Delta y$ is the spatial grid size. The grid size is fixed to $ \Delta
 x/(c/\omega_{pe})=1/40$ and the number of grids in each direction is
 $N_x \times N_y = 20,000 \times 1,680$ throughout this study. The
 relativistic electron cyclotron frequency $\omega_{ce}$ is also the
 characteristic timescale of the relativistic magnetized shocks, and it is
 defined as follows:
 \begin{equation}
  \label{eq:wce}
  \omega_{ce} = \frac{eB_1}{\gamma_1m_ec}.
 \end{equation}

 It is known that in a PIC simulation involving a relativistic plasma
 flow, there appears a so-called numerical Cherenkov instability
 \citep{Godfrey1974}, which will grow to substantial amplitudes and cause
 non-negligible amounts of plasma heating, even in
 homogeneous systems with relativistic bulk flows. As the unphysical
 heating in the cold upstream flow will change the physics of the shock,
 the simulation box size and integration time are severely limited by
 the growth timescale of the numerical instability. To minimize
 numerical artifacts, the method developed by \cite{Ikeya2015} is applied
 to our code. This method is a variant of the method originally reported by
 \cite{Godfrey2013} and \cite{Xu2013}, who found that a particular
 choice of CFL number 
 can dramatically reduce the numerical instability growth rate. However,
 the magic CFL number depends on the numerical scheme used for
 solving Maxwell's equations. \cite{Ikeya2015} performed numerical
 experiments and found that the magic CFL number is equal to unity for
 an implicit Maxwell solver employed in the code. As we used a fixed
 grid size of $\Delta x/ (c/\omega_{pe}) = 1/40$, the time step was automatically
 determined as $\omega_{pe}\Delta t = 1/40$. In addition, we used a
 moving injector for the upstream boundary. The particles are continuously
 injected from an injector that moves away from the shock (i.e., toward
 $+x$ direction) with the speed of light. This minimizes the propagation
 time for the undisturbed upstream uniform plasma, enabling us to follow
 the long-term evolution of the shock without being affected by the
 numerical instability. 

 Note that one has to carefully check numerical artifacts in the code for
 an accurate estimate of the precursor wave emission efficiency. Since
 the precursor waves are high-frequency (and thus short-wavelength)
 electromagnetic waves, they may easily be damped (e.g., in the case of
 the insufficient resolution). In addition, the application of digital
 filtering often used to suppress the numerical Cherenkov noise may
 underestimate the emission efficiency. Therefore, we performed 1D
 simulations and investigated numerical convergence with respect to both
 the grid size and the number of particles (see Appendix 
 \ref{sec:conv}). The numerical resolution employed in the
 2D simulations is motivated by the numerical convergence study. 
  
 We investigated the dependence on the magnetization parameter
 $\sigma_e$ defined as the ratio of the Poynting flux to the kinetic
 energy flux:
 \begin{equation}
  \sigma_e \equiv \frac{B_1^2}{4 \pi \gamma_1 N_1 m_e c^2} =
   \frac{\omega_{ce}^2}{\omega_{pe}^2},  
 \end{equation}
 which is the main controlling parameter of a relativistic magnetized
 shock. Our simulations were performed for the following seven cases:
 $\sigma_e = 1$, $3 \times 10^{-1}$, $1 \times 10^{-1}$, $3\times
 10^{-2}$, $1\times 10^{-2}$, $3\times 10^{-3}$ and $1\times 10^{-3}$.

 \section{Global Shock Structure}\label{sec:shock}
 
  \subsection{High-$\sigma_e$ Case}\label{subsec:highsig}

\begin{figure*}[htb!]
   \plotone{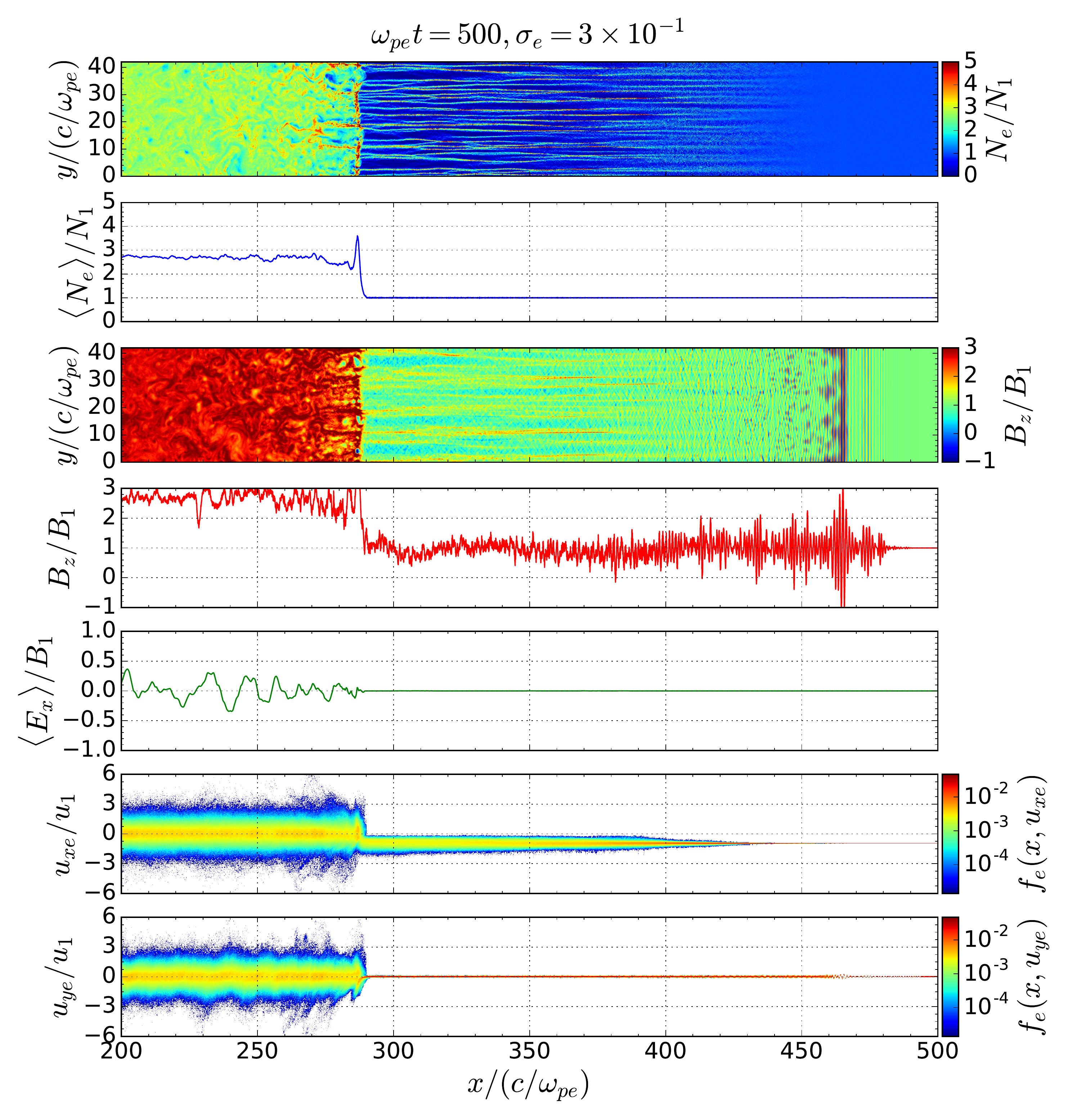}
   \caption{Global shock structures at $\omega_{pe}t=500$ for $\sigma_e
   = 3 \times 10^{-1}$. From top to bottom, the electron number density
   $N_e$, the transversely averaged electron number density $\langle N_e
   \rangle$, the out-of-plane magnetic field $B_z$, the 1D profile for
   $B_z$ along $y/(c/\omega_{pe}) = 21$, the transversely averaged 
   electrostatic field $\langle E_x \rangle$, and the phase-space plots
   of electrons in the $x-u_{xe}$ and $x-u_{ye}$ planes are
   shown. \added{The color scale of the phase-space plots is in a
   logarithmic scale.}}
   \label{out-of-plane_a}
  \end{figure*}

  First, an overview of the global shock structure for a
  relatively high-$\sigma_e$ case is discussed. Figure \ref{out-of-plane_a}
  illustrates the global shock structure at $\omega_{pe} t = 500$ for $\sigma_e =
  3 \times 10^{-1}$. Shown from top to bottom are the electron number
  density $N_e$, the transversely averaged electron number density
  $\langle N_e \rangle$, the out-of-plane magnetic field $B_z$, the 1D
  profile for $B_z$ along $y/(c/\omega_{pe}) = 21$, the transversely averaged 
  electrostatic field $\langle E_x \rangle$, and the phase-space plots 
  of electrons in the $x-u_{xe}$ and $x-u_{ye}$ planes. All quantities 
  are normalized by the corresponding 
  upstream values.  At this time, the shock front is clearly seen at
  $x/(c/\omega_{pe}) \sim 280$. It has already propagated
  sufficiently far away from the left-hand boundary and a well-developed
  shock structure is formed. A shock formation may be identified by the
  density increase by a factor of three, as expected from the MHD
  Rankine-Hugoniot relation (see Appendix 
  \ref{sec:cutoff} for finite deviation from the theoretical prediction). 

  In the upstream region, large-amplitude 
  fluctuations in $B_z$ are clearly seen. These are the electromagnetic
  waves emitted and propagated away from the shock front. The wave 
  magnetic field is polarized in the $z$ direction (i.e., parallel to the
  ambient magnetic field direction). This is the signature of the
  X-mode wave and is consistent with the linear theory of the SMI
  \citep{Hoshino1991} as well as previous 1D PIC  
  simulation results \citep{Langdon1988, Gallant1992, Hoshino1992,
  Amato2006}. At around the tip of the precursor wave region
  ($450 \le x/(c/\omega_{pe}) \le 480$), the wavefront is
  roughly uniform in the $y$ direction and is obviously different from
  the other part of the precursor. The waves in this region are
  generated in the early phase of simulations and are contaminated by
  the initial and boundary conditions. Therefore, we excluded this
  region from our analysis presented below.

  In the precursor region, transverse filamentary structures are clearly
  identified in both the density and magnetic fields. We think that these
  filaments are generated by a filamentation instability discussed by
  \cite{Kaw1973} and \cite{Drake1974} for an ion-electron plasma. This
  is a kind of 
  parametric instability driven by the ponderomotive force of a
  large-amplitude electromagnetic wave. Strictly speaking, to the
  authors' knowledge, this instability has not been analyzed for a pair
  plasma. However, the similarity with the ion-electron plasma case
  indicates that it is probably related to the filamentation
  instability. More comprehensive studies of this instability will be
  presented in a future publication. In any case, the appearance of
  filamentary structures in the precursor region is the strong evidence
  that the precursor waves still retain coherence in 2D systems.
  
  As is seen in the phase-space density plot, the precursor waves cause
  strong heating in $u_x$ in the upstream region. Note that this
  is not clearly seen in $u_y$ because the Lorentz transformation (from
  the upstream rest frame to the laboratory frame) increases the thermal
  spread only in the $u_x$ direction. Because the heating region coincides
  with the density filament region, this may be attributed to
  precursor waves. As discussed in $\S$\ref{subsec:heating}, this apparent
  heating is merely a result of coherent oscillations in velocity in the
  strong wave electromagnetic field and cannot be referred to as a true
  heating.

  The density filaments in the precursor are convected by the upstream
  flow and eventually hit the shock. Consequently, the shock surface is strongly
  modulated and large amplitude fluctuations both in density and
  magnetic field are seen in the downstream region. The downstream
  turbulence \deleted{caused by the interaction between the shock and density
  filaments} may be explained by the Richtmyer-Meshkov instability
  \citep{Richtmyer1960,Meshkov1972} \added{which occurs when a shock
  wave passes through a medium with density inhomogeneity. The density
  filaments interacting with the shock may thus generate turbulence in a
  similar manner.} It is worth noting that the precursor wave activity
  survives even in the presence of such a strong 
  feedback effect of the emission to the shock. As discussed in
  $\S$\ref{subsec:evo}, the wave emission efficiency has nearly
  reached a quasi-stationary state at this stage. This leads to the
  firm conclusion that coherent precursor waves persist in 2D systems.
  
  \subsection{Low-$\sigma_e$ Case}\label{subsec:lowsig}

\begin{figure*}[htb!]
   \plotone{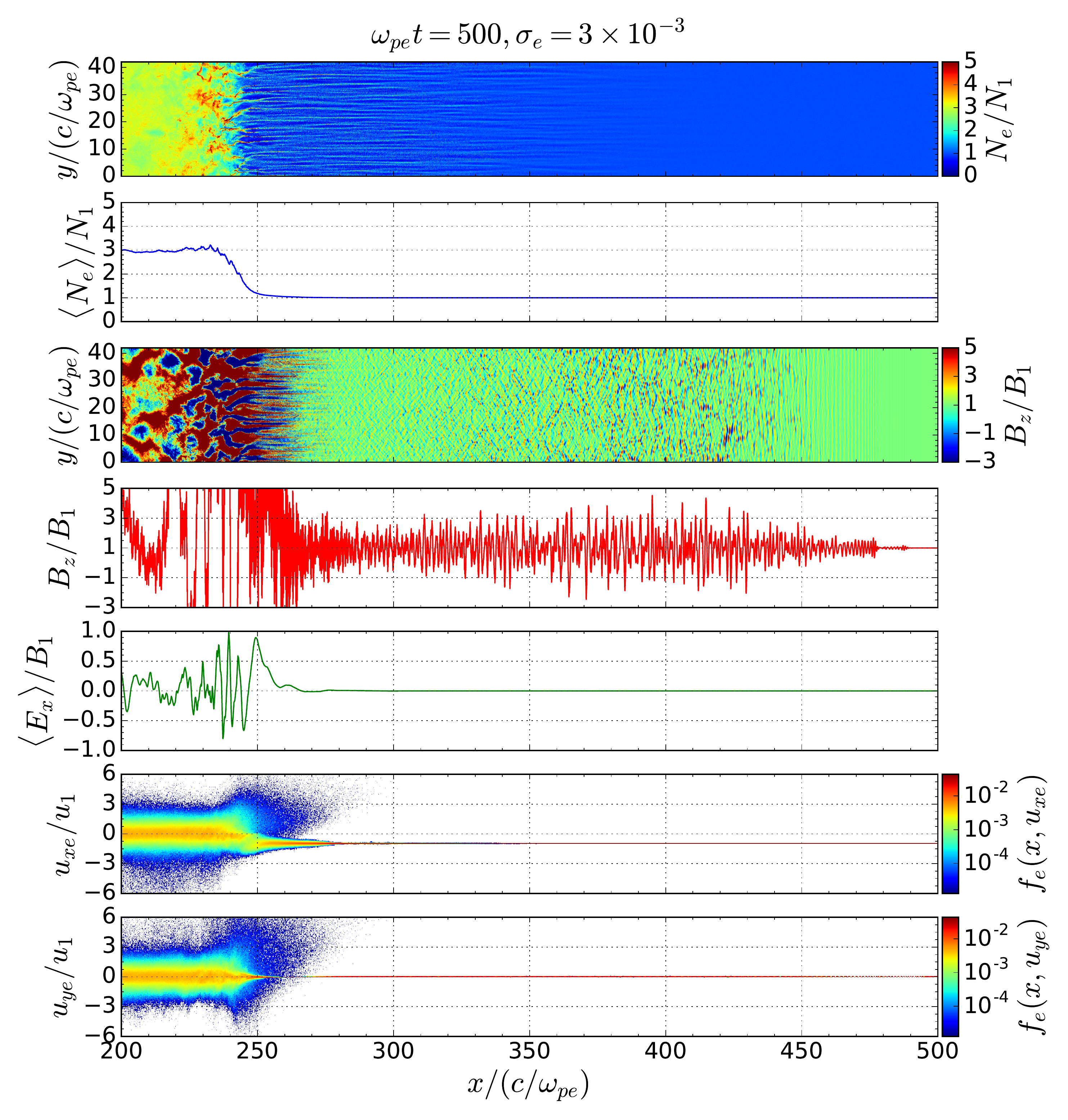}
   \caption{Global shock structures at $\omega_{pe}t=500$ for
   $\sigma_e= 3 \times 10^{-3}$. See the caption of Figure
   \ref{out-of-plane_a} for details. Note that the color scale for $B_z$
   was chosen such that the precursor waves become clearly visible; the
   Weibel-generated magnetic fields in the shock transition region are
   out of scale.}
   \label{out-of-plane_b}
  \end{figure*}

  Further, we discuss the global shock structure for a relatively low
  $\sigma_e$ case. Figure \ref{out-of-plane_b} illustrates the global shock
  structure at $\omega_{pe} t = 500$ for $\sigma_e = 3 \times
  10^{-3}$. The format is the same as Figure \ref{out-of-plane_a}. As we have
  already mentioned, our main concern in a low 
  magnetization regime is the competition between the SMI and the WI. 
  As expected, we observe filamentary density and magnetic field
  structures at $x/(c/\omega_{pe}) = 240 - 280$ in
  Figure \ref{out-of-plane_b}, which are
  attributed to the WI (see Appendix \ref{sec:weibel}). Because the WI results
  from velocity anisotropy near the shock front, the length of
  the magnetic field filament corresponds to that of the reflected
  particle beam (or simply the particle Larmor radius for a perpendicular
  shock). It is clear that the density filaments are extended well ahead
  of the Weibel-dominated shock transition region where no
  back-streaming particles exist. We believe that these filaments are
  generated by the interaction between coherent radiation and
  upstream plasmas, just as in the high-$\sigma_e$ case discussed in
  $\S$\ref{subsec:highsig} .

  Surprisingly, precursor waves are observed with appreciable
  amplitudes in this case as well. Although
  the amplitude of the density filaments is smaller than that in the
  high-$\sigma_e$ case, presumably due to smaller precursor wave
  amplitudes (see $\S$\ref{subsec:dependence}), the formation of
  filaments indicates  
  that the precursors strongly interact with the upstream plasma. Particle
  heating in the precursor region is not observed in Figure
  \ref{out-of-plane_b} again because 
  of lower wave amplitude. Nevertheless, finite heating due to quiver
  motion in the wave electromagnetic field is indeed identified (see
  $\S$\ref{subsec:heating}). All these results suggest that the
  precursor waves remain coherent even in this case, and that the SMI
  and the WI may somehow coexist at a relativistic magnetized shock.

 \section{Precursor wave properties}\label{sec:precursor}
 
  \subsection{Time Evolution}\label{subsec:evo}

\begin{figure*}[htb!]
   \plottwo{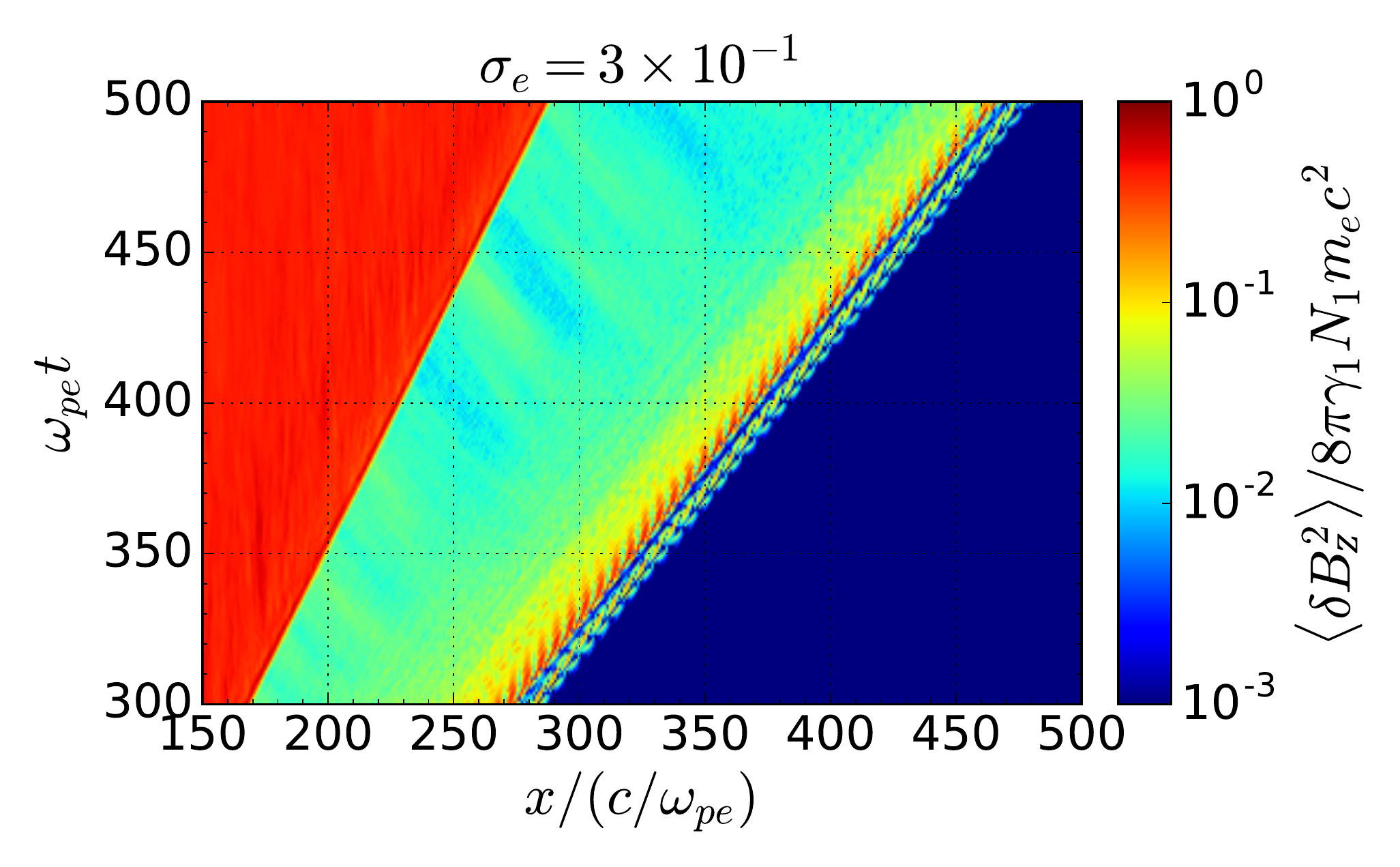}{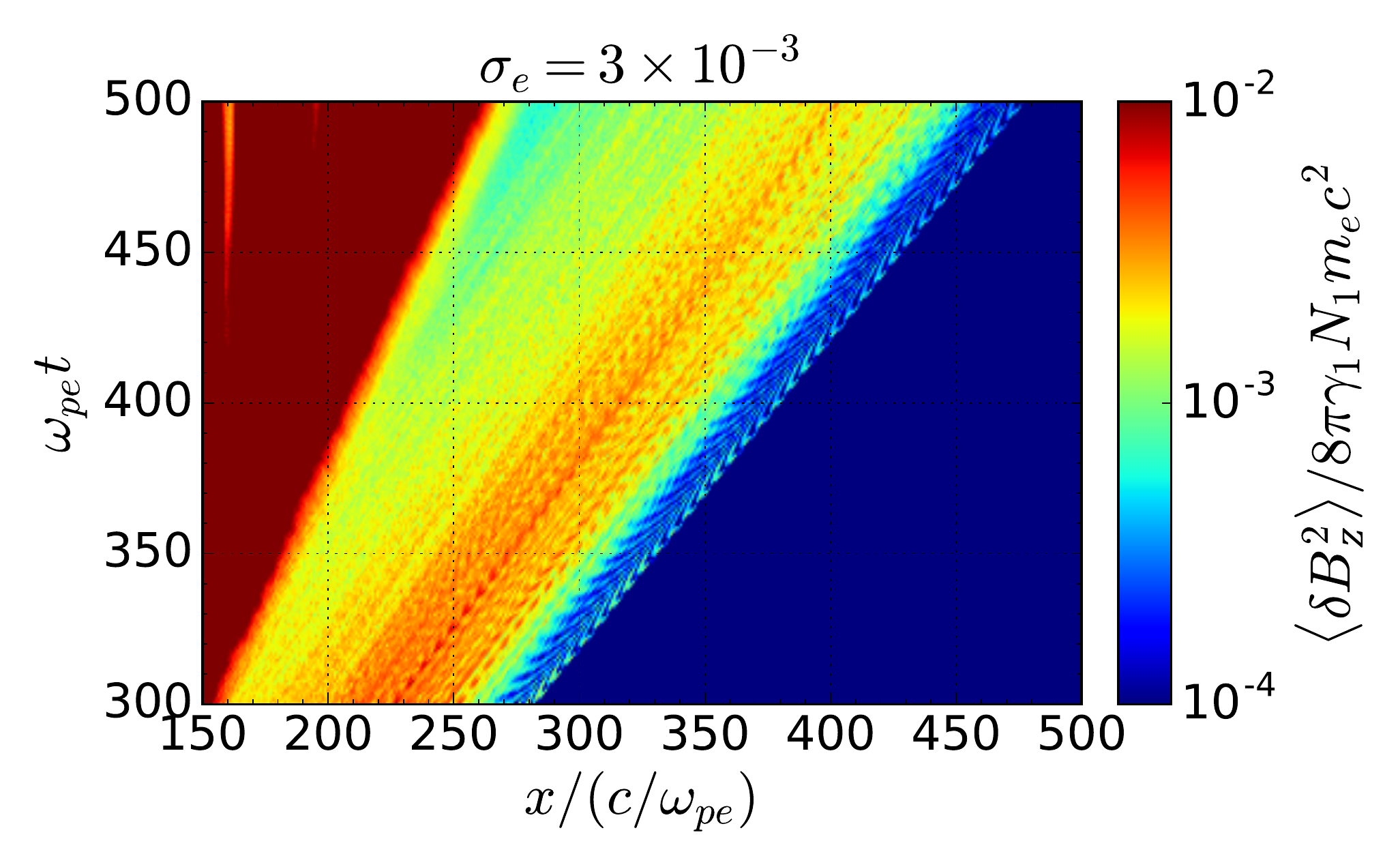}
   \caption{Time evolution of the wave energy averaged over the $y$
   direction given in units of upstream bulk kinetic energy:
   $\sigma_e=3\times 10^{-1}$ (left) and $\sigma_e=3 \times 10^{-3}$
   (right).}
   \label{xt}
  \end{figure*}

  Here, we discuss the time evolution of precursor wave power. Figure
  \ref{xt} shows the time evolution of the wave energy averaged over the $y$ 
  direction  \added{$\langle \delta B_z \rangle / 8\pi \gamma_1 N_1 m_e
  c^2$ (where $\delta B_z \equiv B_z - B_1$ is the fluctuating magnetic
  field component)} in the range $300 \le \omega_{pe}t \le 500$. The
  wave energy is  
  normalized by the upstream bulk kinetic energy. Two different runs with
  $\sigma_e = 3 \times 10^{-1}$ (left) and $\sigma_e = 3 \times 10^{-3}$
  (right) are shown. \added{Note that since the precursor waves are
  high-frequency electromagnetic waves, the same plots for $\delta E_y$
  look almost identical. } 

  This study primarily focuses on a sufficiently long-term evolution
  wherein the effects of the initial and boundary conditions may be 
  neglected. For this purpose, we focus on precursor waves in
  the immediate upstream of the shock front, as they are newly generated
  waves at the shock in the highly-disturbed medium. Figure \ref{xt}
  indicates that the precursor wave amplitude 
  near the shock front gradually decreases over time. However, it remains
  finite and reaches a quasi-steady state by the end of the simulations.

\begin{figure*}[htb!]
   \plottwo{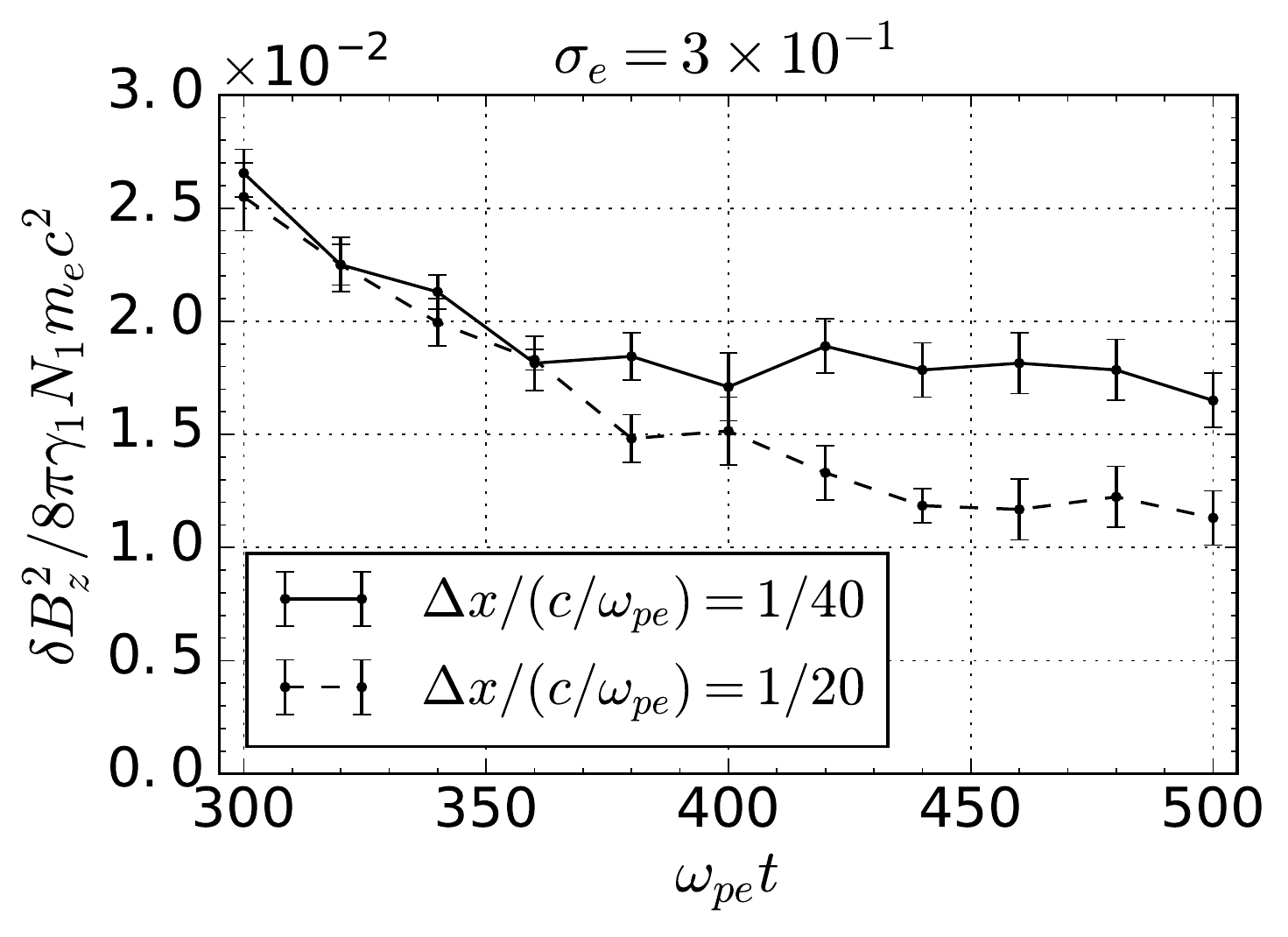}{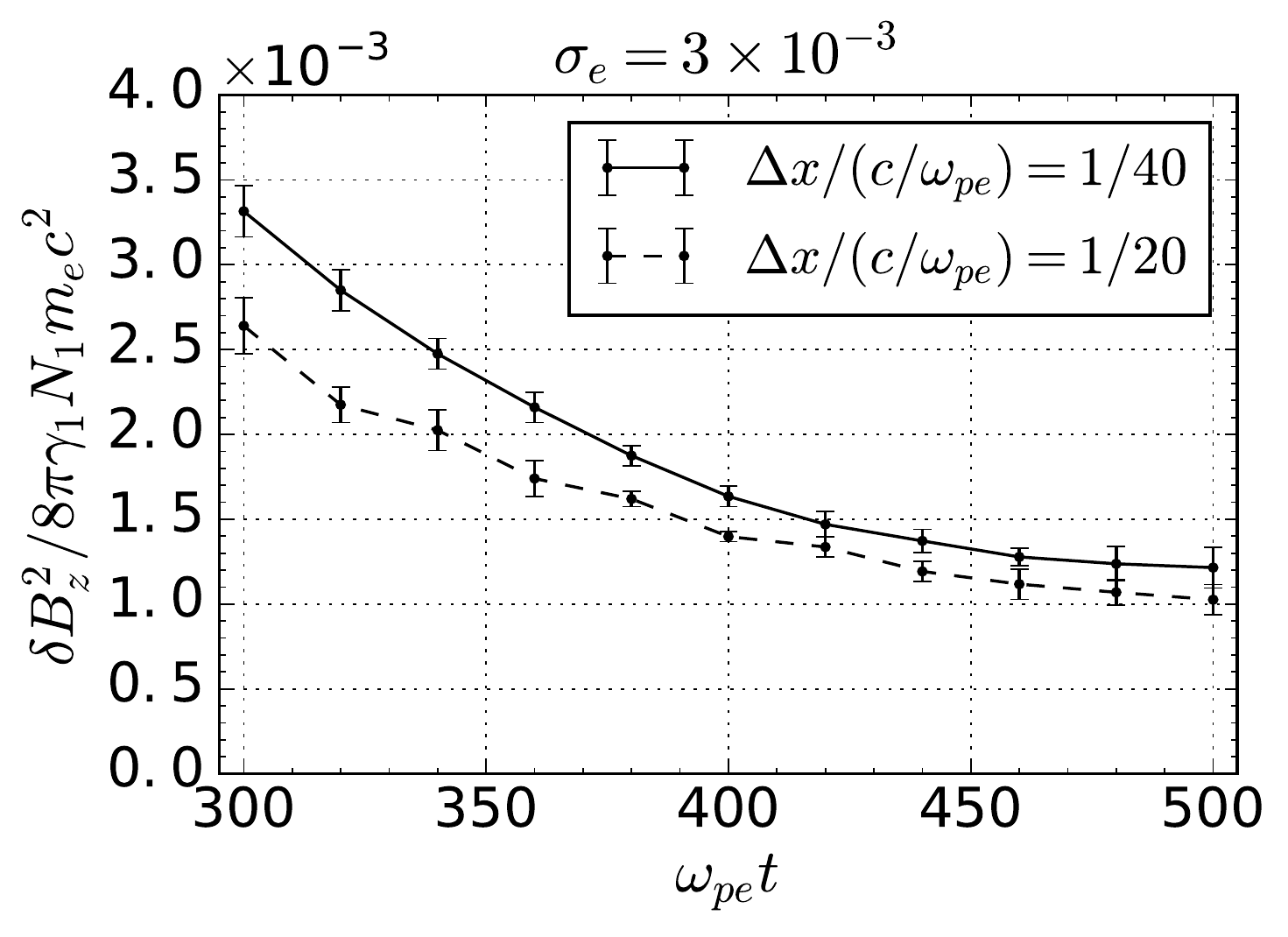}
   \caption{Time evolution of the precursor wave energy given in units of
   upstream bulk kinetic energy: $\sigma_e = 3 \times 10^{-1}$ (left)
   and $\sigma_e = 3 \times 10^{-3}$ (right). In both panels, the
   results with our fiducial resolution ($\Delta x/ (c/\omega_{pe})
   =1/40$) are shown by the solid lines, whereas lower resolution
   results ($ \Delta x/(c/\omega_{pe}) = 1/20$) are shown by the dashed
   lines.}
  \label{evolution}
  \end{figure*}

  We quantified the precursor wave power by taking the average
  power in the range: $r_L < x - X_{sh} < r_L + 50 c/\omega_{pe}$, where
  $X_{sh}$ is the position of the shock and $r_L = c/\omega_{ce}$ is the
  relativistic Larmor radius calculated with the upstream flow Lorentz
  factor $\gamma_1$. The shock position $X_{sh}$ is determined from
  the $y$-averaged density profile $\langle N_e \rangle$, assuming that
  the shock propagation speed is constant over time. Figure 
  \ref{evolution} shows the precursor 
  wave energy density as a function of time in the range $ 300 \le
  \omega_{pe}t \le 500$ for $\sigma_e = 3 \times 10^{-1}$ (left) and $3 \times
  10^{-3}$ (right). Note that the results with low resolution $\Delta
  x/(c/\omega_{pe}) =1/20$ are also shown for comparison. Each data point
  represents the temporal average during the time interval $20/\omega_{pe}$,
  and the error bars were estimated from the standard deviation during
  the time interval. The error bars for $\sigma_e=3\times 10^{-1}$ were
  systematically larger than those for $\sigma_e=3\times 10^{-3}$ because of
  low-frequency fluctuations in $B_z$ associated with the density
  filaments. Nevertheless, the results suggest that the precursor wave
  power is nearly saturated at around $\omega_{pe} t \simeq 450$ in both
  cases at our fiducial resolution.
    
  On the other hand, the lower resolution runs showed continuous decrease
  (for $\sigma_e = 3 \times 10^{-3}$) or saturation at a lower level (for
  $\sigma_e = 3 \times 10^{-1}$). Therefore, it is natural that even
  lower resolution simulations reported previously could not accurately
  reproduce the precursor waves \citep{Sironi2009, Sironi2011}. 

  \subsection{Precursor Heating}\label{subsec:heating}

  As shown in Figure \ref{out-of-plane_a}, the $x$
  component of the electron four-velocity $u_{xe}$ in the precursor
  wave region for $\sigma_e = 3 \times 10^{-1}$ has a substantial
  spread. However, we found that 
  particles' velocities are merely oscillating in the precursor wave
  electromagnetic field and that the apparent broadening in the velocity
  distribution does not indicate true heating. 

\begin{figure*}[htb!]
   \plotone{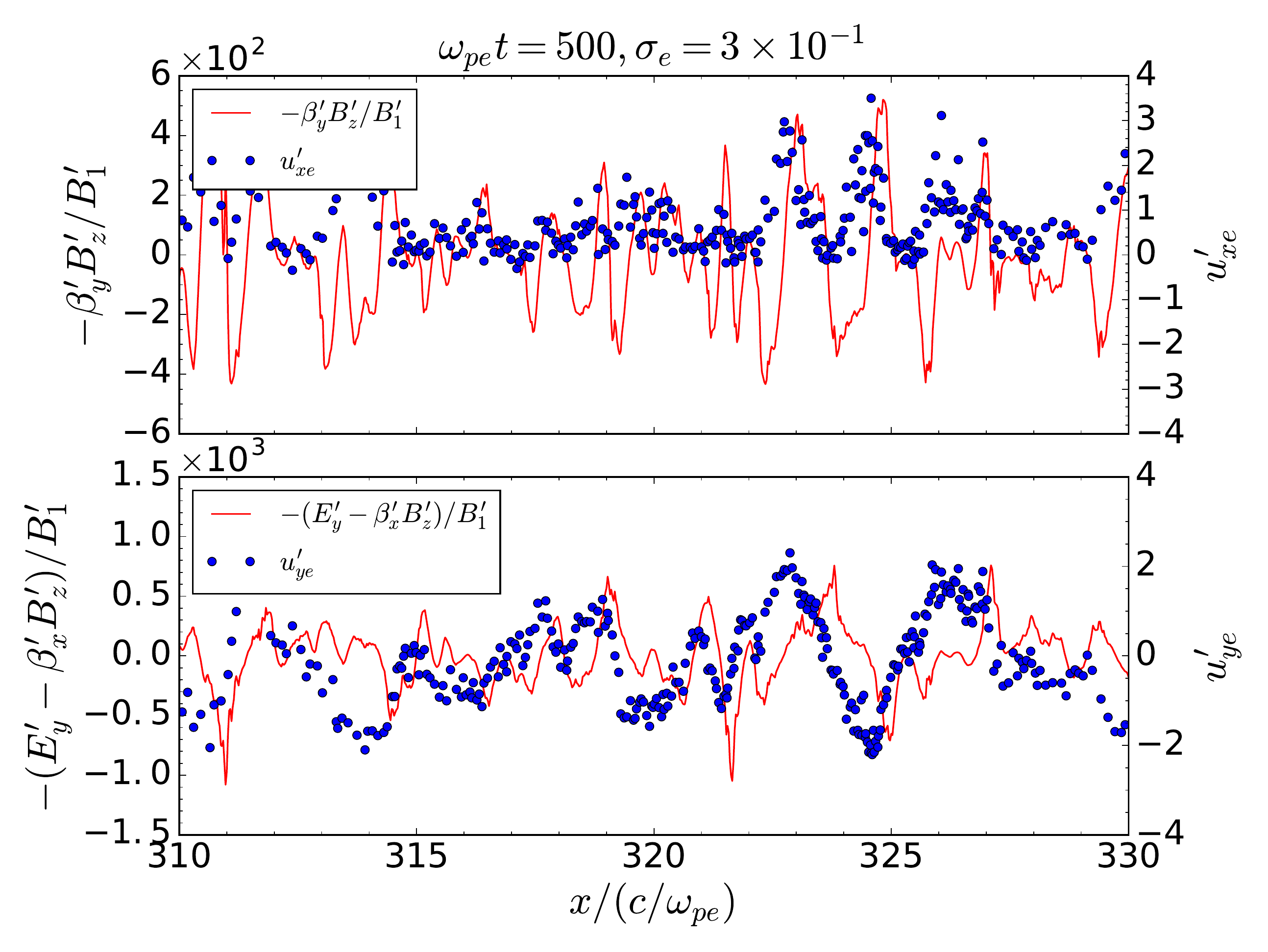}
   \caption{Comparison of the Lorentz force (red solid line) with individual
   electron velocities (blue circle) for $\sigma_e = 3 \times
   10^{-1}$. The top and bottom panels show the $x$ and $y$ components,
   respectively.} 
   \label{xu-high}
  \end{figure*}

\begin{figure*}[htb!]
   \plotone{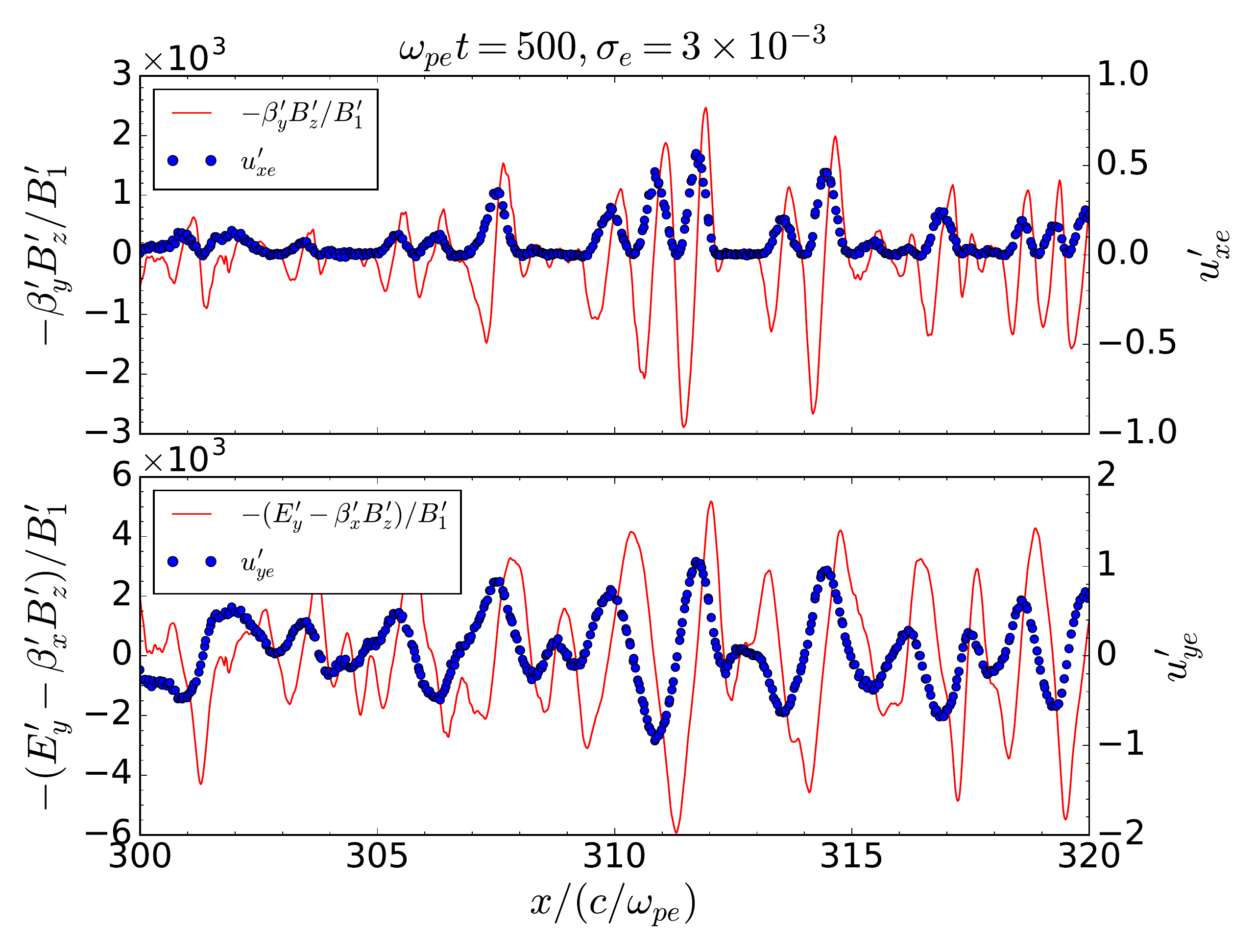}
   \caption{Comparison of the Lorentz force with individual electron
   velocities for $\sigma_e = 3 \times 10^{-3}$. See the caption of Figure
   \ref{xu-high} for details.}
   \label{xu-low}
  \end{figure*}

  To confirm this idea, Figures \ref{xu-high} and \ref{xu-low} show
  individual electron four-velocities (blue circle) and the Lorentz
  force (red solid line) \added{both in the upstream rest frame}
  calculated at each grid point along 
  $y/(c/\omega_{pe})=21$ for $\sigma_e = 3 \times 10^{-1}$ and $\sigma_e =
  3 \times 10^{-3}$, respectively. Note that $E_x$ is weak because of the
  perfect symmetry in a pair plasma. The top and 
  bottom panels show the $x$ and $y$ components, respectively, at
  $\omega_{pe}t = 500$. In addition, only ~1\% of the electrons located
  within one grid cell in the $y$ direction are 
  shown. \added{The prime indicates physical quantities measured in the
  upstream rest frame.} \added{It is known that an electron motion in a plane,
  linearly polarized monochromatic electromagnetic wave (with $\omega =
  k c$) propagating in the $+x$ direction may be written as 
  \begin{eqnarray}
   \label{eq:ux}
   u_x &=& \frac{a^2}{2}\cos^2{\omega\left(\frac{x}{c}-t\right)}, \\
   \label{eq:uy}
   u_y &=& a\cos{\omega\left(\frac{x}{c}-t\right)},
  \end{eqnarray}
  where $a =e\delta E/m_ec\omega$ is the strength parameter
  \citep[see][]{Lyubarsky2006} . }  
  \replaced{The electron quiver motion is 
  correlated quite well with the Lorentz force exerted by the precursor
  waves}{The electrons obey these equations quite well}, suggesting that
  the particles in the upstream are not actually 
  thermalized but forced to oscillate in a coherent manner by the wave
  electromagnetic field. 
 
  It is easy to confirm that the electromagnetic waves propagating in a
  plasma always have superluminal phase speeds. Therefore, they are not
  susceptible to resonant wave-particle interactions and will not
  directly thermalize the particles. Heating may occur when a
  parametric instability produces longitudinal waves with subluminal
  phase speeds, allowing energy to be absorbed by the particles
  via wave-particle resonances. The WFA is indeed one of the
  examples of such heating. However, in the present simulations,
  although filamentation instability occurs, the electric field
  associated with the density filaments is weak due to the symmetry in a
  pair plasma. This is probably the reason why there is no true heating
  in the precursor even in the presence of filamentation instability.
  
  In any case, the fast particle quiver motion may behave as an
  effective temperature that potentially reduces the growth rate of the
  SMI.  Note that the SMI results from harmonic
  resonances between the relativistic particle gyromotion and the X-mode
  wave $\omega = n\omega_{ce}$, where integer $n$ denotes the
  harmonic number of the resonance.
  \cite{Amato2006} reported that a finite thermal 
  spread in the ring distribution suppresses the growth rate at higher
  harmonics, and the suppression becomes
  increasingly significant at lower harmonic numbers as the thermal spread of the
  particles increases. Therefore, we think that the effective
  temperature may be responsible for the gradual decrease of the
  precursor wave amplitude. The quasi-steady state in the final phase
  of the simulations is probably determined by the balance between wave
  generation and heating.

  \subsection{Wavenumber Spectra}\label{subsec:spectra}

\begin{figure*}[htb!]
   \plottwo{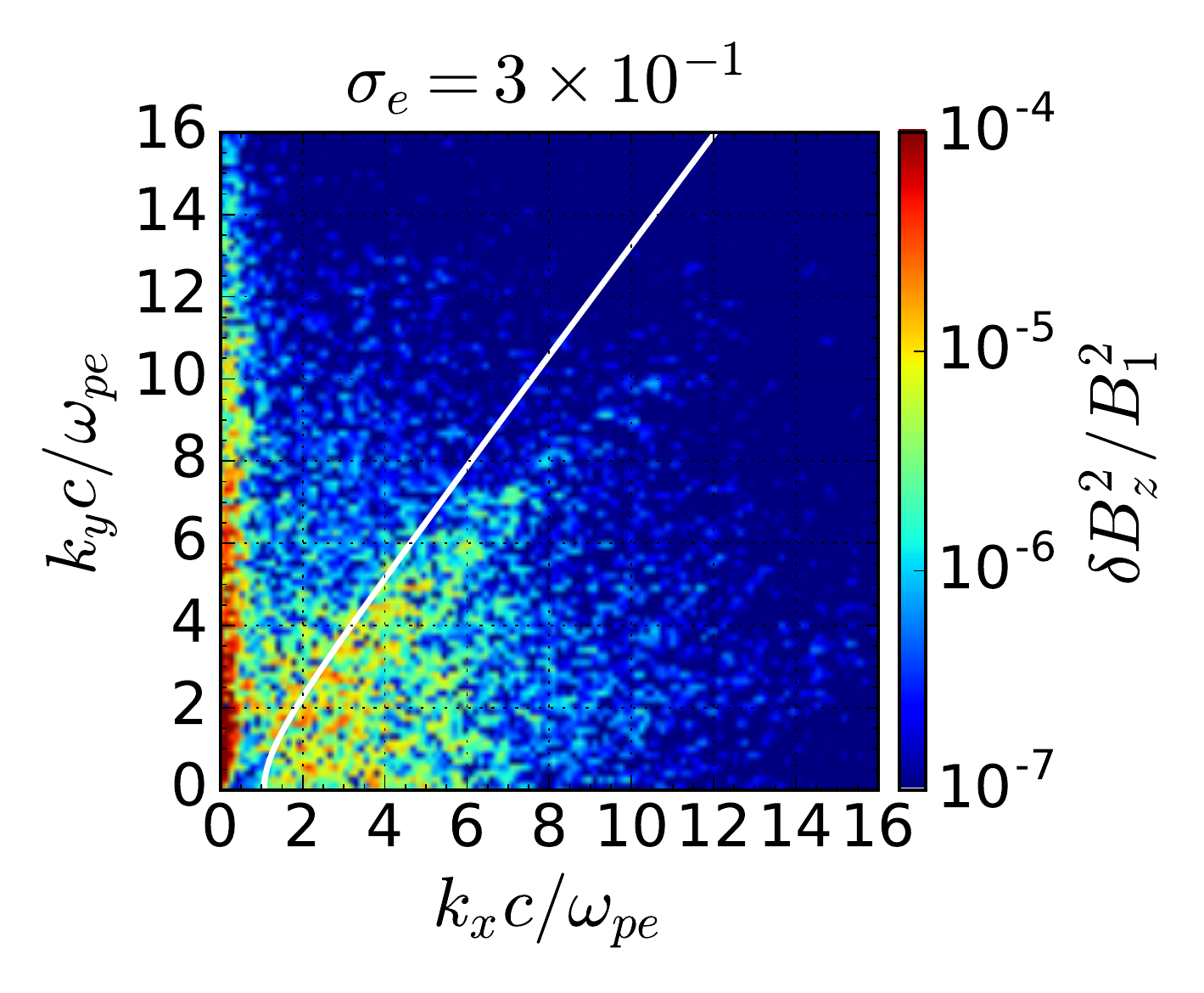}{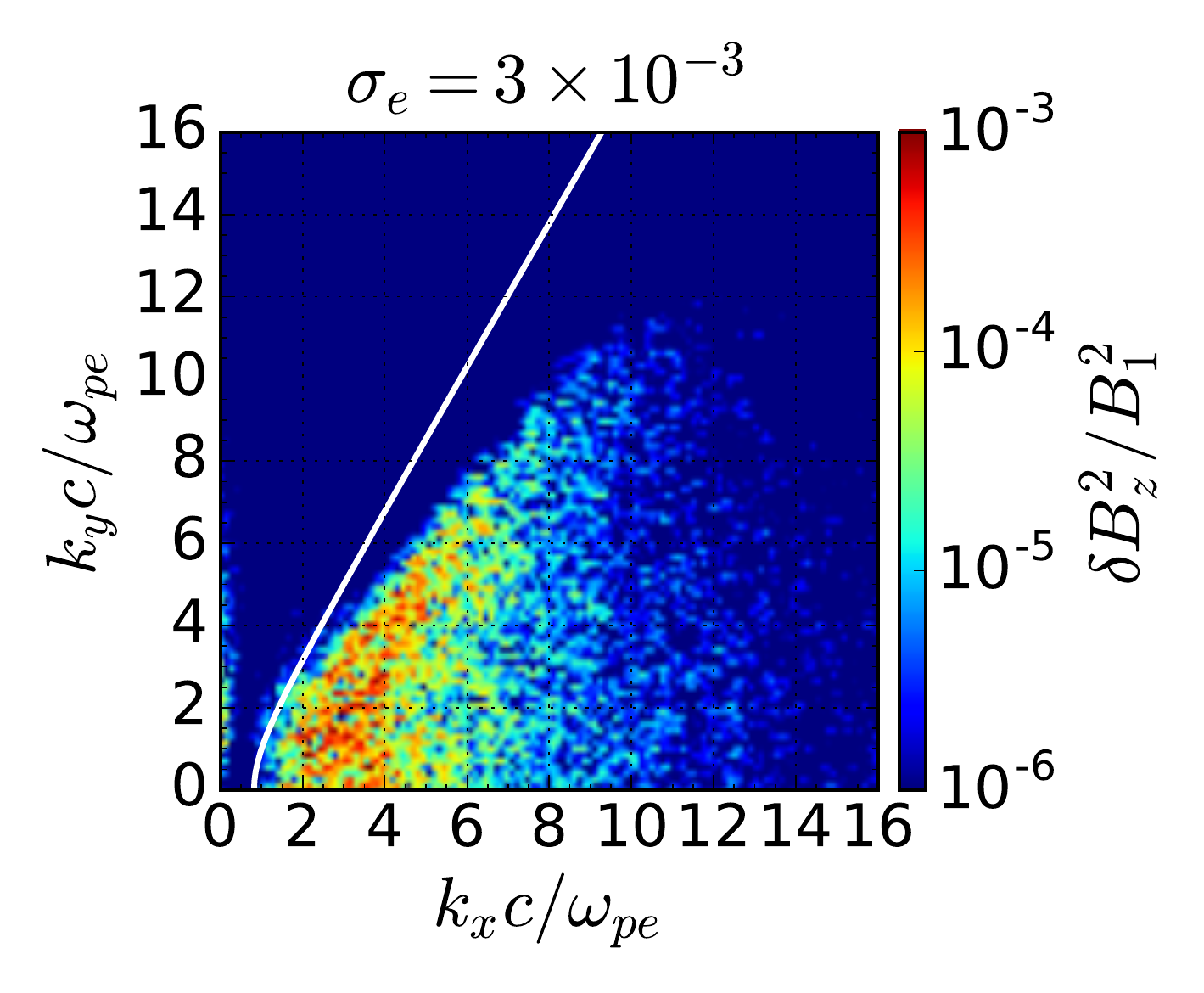}
   \caption{Wavenumber power spectra for a precursor wave at
   $\omega_{pe}t=500$ for $\sigma_e = 3 \times10^{-1}$ (left) and
   $\sigma_e = 3 \times10^{-3}$ (right). The white solid line indicates
   a theoretical cutoff wavenumber (see Appendix \ref{sec:cutoff}).}  
   \label{disp}
  \end{figure*}

  Figure \ref{disp} shows the precursor wave power spectra in wavenumber
  space normalized by the upstream ambient magnetic field energy density.
  The spectra were obtained for a snapshot at $\omega_{pe}t =500$ by
  taking the Fourier transform in the region: $r_L < x - X_{sh} < r_L +
  50 c/\omega_{pe}$. We applied the Hanning window to remove edge
  effects. Note that the Nyquist wavenumber for our simulation is $k_N =
  \pi/\Delta x \sim 120\omega_{pe}/c$, and the precursor waves are well
  resolved.

  Each of these spectra has a lower cutoff wavenumber. This can be
  explained by the wavenumber above which the upstream-directed group
  velocity of the precursor wave is greater than the shock propagation
  velocity \citep{Gallant1992}. We estimated the theoretical cutoff
  wavenumber (see Appendix \ref{sec:cutoff}), which is shown by the white
  solid line in Figure  \ref{disp}. The theoretical lower cutoff
  wavenumber is roughly 
  consistent with the measured one. The wave power to the right of the
  white solid line may be attributed to the precursor waves propagating
  away from the shock front. Note that both spectra have 
  non-negligible wave power around $k_x =0$, and in particular the power for
  $\sigma_e = 3 \times 10^{-1}$ has a substantial fraction of the total wave
  power. Recall that we excluded the region contaminated by the WI from
  this analysis. Therefore, we think that waves around $k_x =0$ are
  associated with the filamentation instability. These filamentary
  structures generated in the upstream region are convected by the
  upstream plasma flow, which then interact with the shock.

  \subsection{$\sigma_e$ Dependence}\label{subsec:dependence}
  
\begin{figure*}[htb!]
   \plottwo{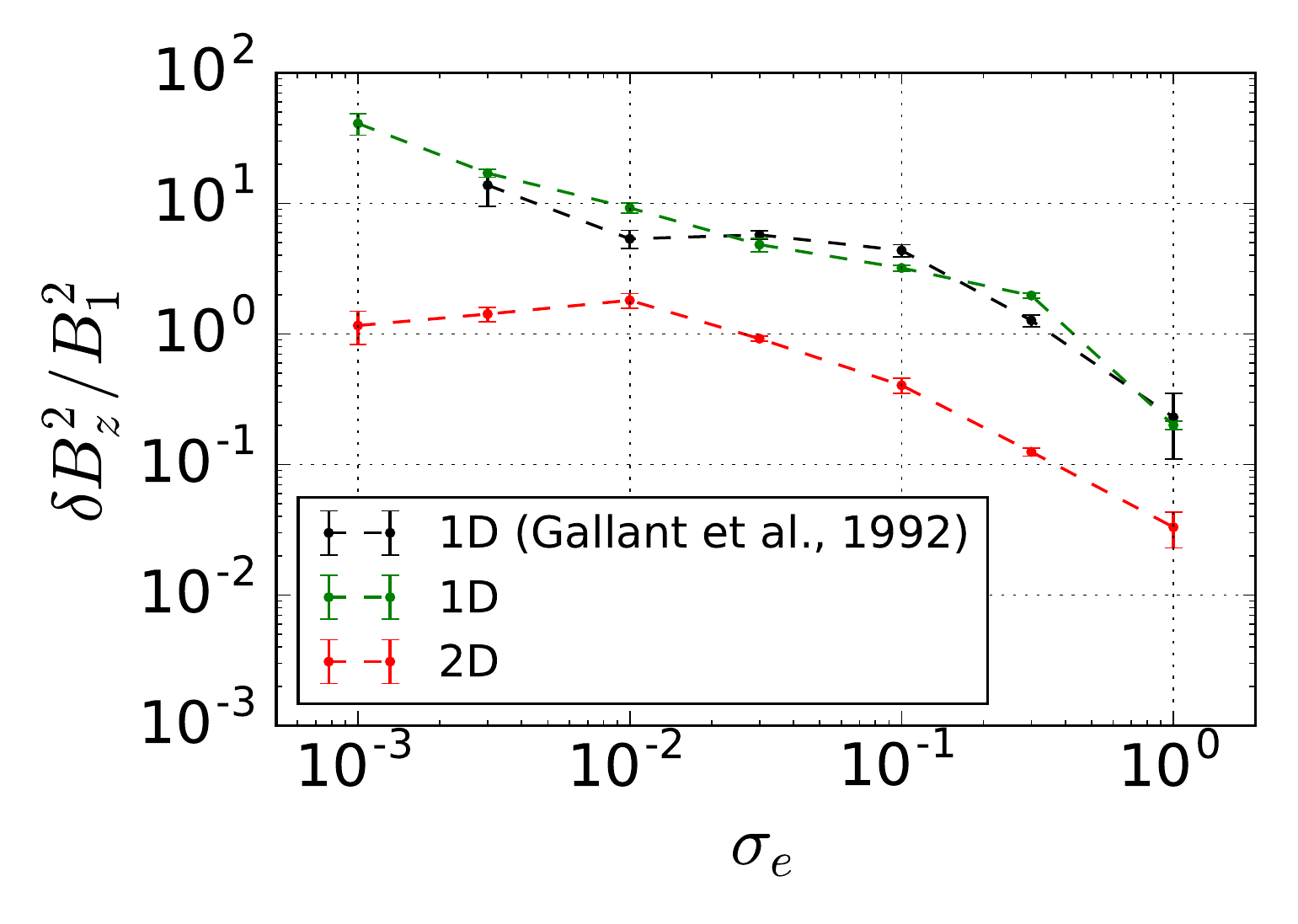}{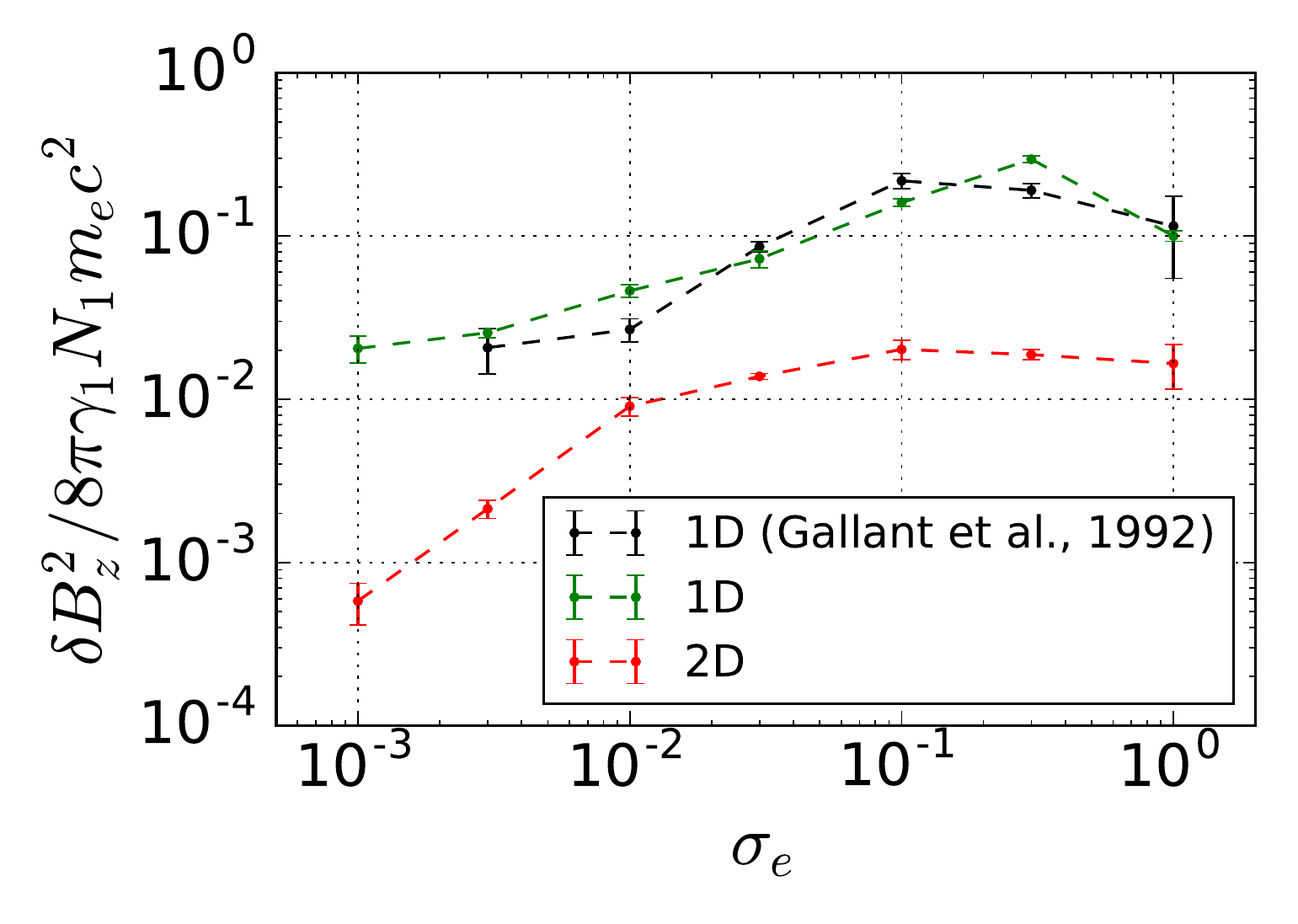}
   \caption{$\sigma_e$ dependence of the precursor wave
   energy in two different normalizations. Left: normalized to the
   ambient upstream magnetic field energy. Right: normalized to the
   upstream kinetic energy. The 1D simulation results
   by \cite{Gallant1992} as well as the 1D and 2D simulation results of
   this study are shown in black, green and red, respectively.}
   \label{amp}
  \end{figure*}

  Now we discuss the $\sigma_e$ dependence of the precursor wave
  amplitude. The amplitude was calculated by integrating the power 
  spectra as described above over the whole wavenumber space. Figure
  \ref{amp} shows the precursor wave power as a function 
  of $\sigma_e$ with two different normalizations: one normalized to the
  upstream ambient magnetic field (left) and another to the upstream bulk
  kinetic energy (right). Figure \ref{amp} also shows 1D
  simulation results and the results reported by \cite{Gallant1992}
  for comparison. The amplitudes of the precursor waves in 2D were
  systematically smaller than those in the 1D results. This is more or
  less to be expected. Because the inhomogeneity along the shock surface
  reduces the coherence of shock-reflected particles, the cold ring
  distribution assumption may no longer valid and the
  growth rate and saturation level of the SMI may become
  smaller than that in 1D simulations. The $\sigma_e$ 
  dependence shows a different behavior for $\sigma_e \le
  10^{-2}$, which may be attributed to the WI. Although the wave power
  was smaller than that in 1D results by roughly an order of magnitude in the
  Weibel-dominated regime, the coherent precursor waves continued to
  exist with amplitudes sufficiently strong to substantially disturb the
  upstream medium, confirming that the coherent electromagnetic
  precursor wave emission is indeed an intrinsic aspect of relativistic
  shocks. 

 \section{Discussion}\label{sec:discussion}
  
  \subsection{SMI in the Weibel Region}\label{subsec:smi} 

  Our simulation results suggest that precursor wave emission via the
  SMI occurs even in the regime where the WI produces a large-amplitude
  magnetic turbulence. This sounds rather counterintuitive in the sense
  that the WI and the SMI both grow from the same unstable
  distribution function: the dominant WI in a low-$\sigma_e$
  shock may substantially distort the distribution function before
  electromagnetic waves are amplified by the SMI. However, in reality,
  this is not the case and both instabilities may actually
  coexist to some extent. This may be understood as follows.
  
  We may roughly estimate the effect of the Weibel-generated turbulence
  on the particle trajectory by comparing the Lorentz force exerted by
  the ambient $B_1$ and the fluctuating magnetic field components
  $\delta B$. For this, we assume that the fluctuation is random at
  scale length longer than the coherence length $\lambda$ of the
  turbulence. By assumption, the turbulent magnetic field effect
  disappears if the time scale is longer than the particle transit time
  over the coherence length $\lambda /c$. Therefore, if the Lorentz
  force arising from the fluctuation during this time scale $\propto
  \delta B \lambda /c$ is smaller than the average Lorentz force over the
  unperturbed cyclotron period $B_1/\omega_{ce}$:
  \begin{equation}
   \label{eq:smi1}
    \delta B \frac{\lambda}{c} \lesssim \frac{B_1}{\omega_{ce}},
  \end{equation}
  the particle performs, on average, a gyromotion with respect to the
  ambient magnetic 
  field. This condition (Eq. \ref{eq:smi1}) may be written as follows:
  \begin{equation}
   \label{eq:smi2}
   \epsilon_B \lesssim \left( \frac{\lambda}{c/\omega_{pe}}\right)^{-2},
  \end{equation}
  where $\epsilon_B = \delta B^2/ 4\pi N_1\gamma_1 m_e c^2$ is the
  energy conversion rate from upstream plasma kinetic energy into
  Weibel-generated magnetic field energy. Equivalently, this may be
  understood as the condition such that the Larmor radius defined with
  respect to the fluctuation magnetic field is larger than the coherence
  length of the turbulence. Therefore, the particle cannot complete a
  full gyromotion around the turbulent magnetic field.

  It is interesting to note that the above condition is solely
  determined by the properties of the WI, and independent of the shock
  parameters such as $\sigma_e$ or $\gamma_1$. Noting
  that our simulations give $\epsilon_B \sim O(0.1)$ and $\lambda/(c/\omega_{pe})
  \sim O(1) $, we find that the condition (Eq. \ref{eq:smi2}) is
  always satisfied, and the unperturbed particle gyromotion is sustained. We 
  believe that this is the reason why we observe large-amplitude waves
  even in the presence of the strong Weibel turbulence.

  Nevertheless, the effect of turbulence may not completely be
  neglected. Since the turbulent magnetic field fluctuation gives random
  kicks during the gyromotion, the coherence of the particle motion may
  be broken. In other words, the fluctuation introduces an effective
  thermal spread in the original otherwise cold ring distribution in
  momentum. As we already mentioned in $\S$\ref{subsec:heating}, a
  finite temperature reduces the growth rate of the SMI at higher
  harmonics \citep{Amato2006}. Therefore, the wave power at higher
  frequency will be strongly reduced in the Weibel turbulence. This
  probably leads to the lower wave emission efficiency in the
  Weibel-dominated regime, but the lower frequency waves being less
  affected by the turbulence may persist with finite amplitude. 

  On the other hand, we expect that there will be a lower limit for the
  magnetization rate below which coherent precursor waves cannot be
  generated. A lower magnetization rate requires a higher 
  harmonic number $n$ for a generated wave to propagate upstream because the 
  wave frequency must be larger than the cutoff frequency $\omega = n
  \omega_{ce} \gtrsim \sqrt{2(1+\beta_{shock}\gamma_{shock})}
  \omega_{pe}$ (see Appendix \ref{sec:cutoff}). Therefore,  
  the suppression of higher harmonic resonance due to the finite
  temperature effect becomes progressively important as decreasing
  $\sigma_e$. More quantitative estimate for the lower limit will be
  addressed in the future.


  \subsection{Particle Energy Spectra}\label{subsec:energy}

\begin{figure*}[htb!]
   \plottwo{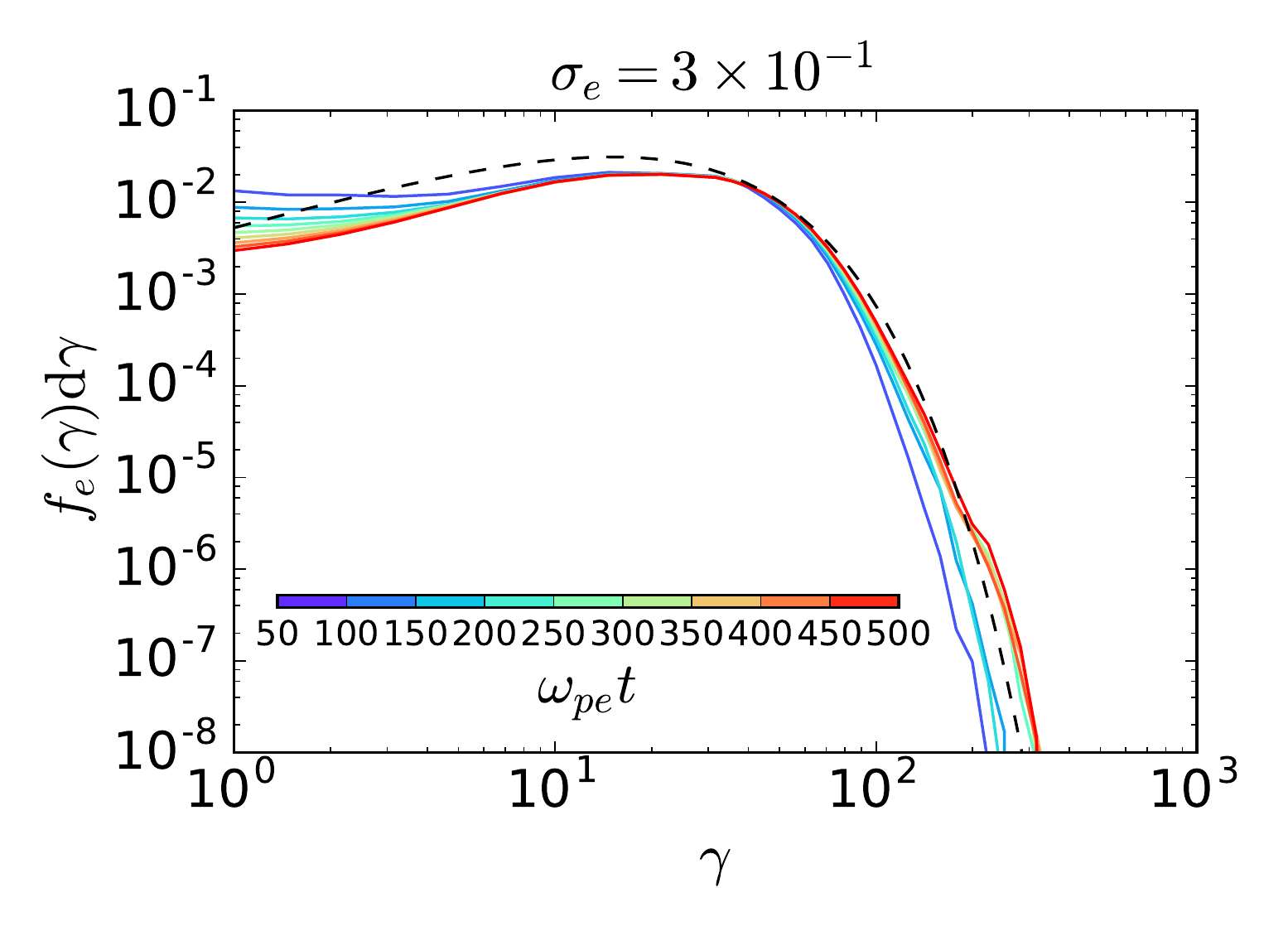}{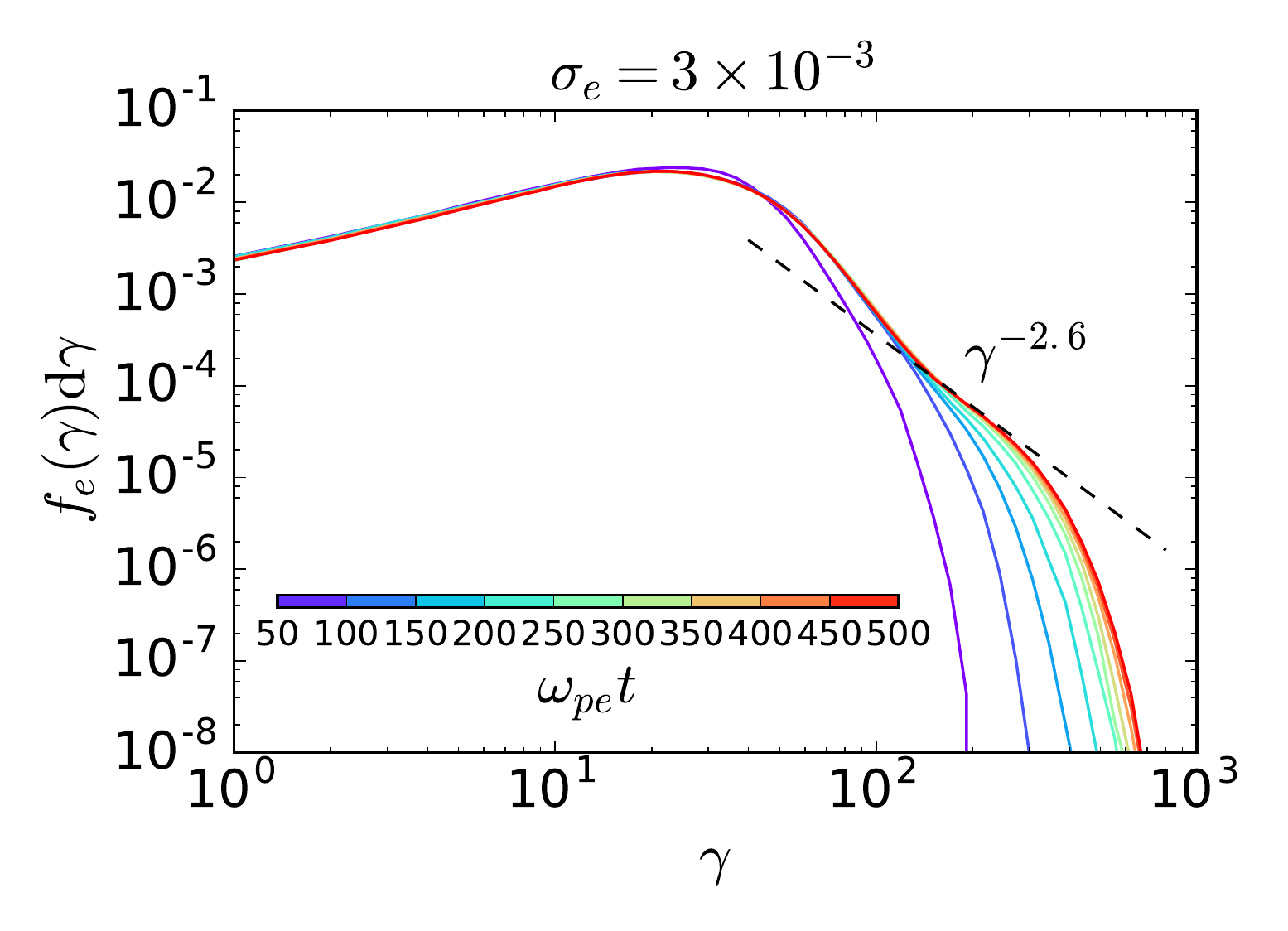}
   \caption{Downstream energy spectra of electrons: $\sigma_e = 3
   \times 10^{-1}$ (left) and $\sigma_e = 3 \times 10^{-3}$ (right). The
   black dashed lines in the left and right panels indicate a 2D 
   relativistic Maxwellian and a power law distribution fitting result,
   respectively. }
  \label{distribution}
  \end{figure*}

  Figure \ref{distribution} shows the downstream energy spectra of
  electrons for $\sigma_e = 3 \times 10^{-1}$ and $3 \times 10^{-3}$ in
  the range $r_L < x < X_{sh}-r_L$, which are normalized as follows:
  \begin{equation}
    \int^{\infty}_1 f_e({\gamma}) {\rm d}\gamma = 1.
  \end{equation} 
  The positron energy spectra are identical to those of electrons. The
  time evolution from $\omega_{pe}t = 50$ to $\omega_{pe}t = 500$ is
  shown. For $\sigma_e = 3 \times 10^{-1}$, the   
  measured distribution reaches a steady state at $\omega_{pe}t = 400$
  and can be well fitted with the 2D relativistic Maxwellian, 
  \begin{equation}
   f(\gamma){\rm d}\gamma \propto \gamma \exp(-\frac{\gamma m c^2}{kT}).
  \end{equation}
  This indicates that the particles downstream are
  completely thermalized and that the non-thermal particles are not
  generated. In contrast, for
  $\sigma_e = 3 \times 10^{-3}$, a suprathermal tail is clearly
  visible. The tail gradually approaches a power law with a spectral 
  index of $p \sim 2.6$. The energy spectrum reaches a steady state at
  $\omega_{pe}t \sim 450$, and the maximum Lorentz factor was approximately at
  $\gamma_{sat} \sim 600$. \cite{Sironi2013} reported that
  efficient particle acceleration 
  occurs in a relativistic perpendicular shock propagating in a pair
  plasma due to the strong turbulence generated by the WI when $\sigma_e
  \lesssim 3 \times 10^{-3}$, and that the acceleration efficiency is
  independent of the bulk Lorentz factor $\gamma_1$ if $\gamma_1
  \gtrsim 10$. Note that $\gamma_1 = 40$ in our simulations satisfies
  this condition. However, the energy spectrum in our
  simulation saturated faster than that in their simulation possibly
  because our high-resolution simulation without the application of
  digital filtering can accurately resolve Weibel-generated turbulence.
  So far, we have not found any evidence of precursor wave emission
  contributing positively for the particle acceleration in a pair plasma
  shock.  

  \subsection{Applicability of WFA}\label{subsec:wfa}

  Based on the simulation results, we now discuss the application of the
  WFA model to astrophysical relativistic shocks. As mentioned earlier,
  \cite{Lyubarsky2006} and \cite{Hoshino2008} both presented
  1D simulation results. In contrast, the numerical
  resolution of the 2D simulations presented by \cite{Sironi2011} was
  probably insufficient for this purpose. However, another approach was taken
  by \cite{Kuramitsu2008}, who assumed the presence of
  precursor waves in their 2D PIC simulations. Focusing only on the
  interaction between the precursor and plasma in the upstream
  region, they injected large-amplitude electromagnetic waves into a
  uniform ion-electron plasma and investigated particle acceleration
  efficiency. Although based on the strong assumption, they found that
  efficient particle acceleration indeed occurs when the strength
  parameter of the electromagnetic wave $a = e\delta E/m_ec\omega$ is
  greater than unity,  where $\delta E$ is the amplitude of the wave
  electric field, and $\omega$ is the wave frequency. The accelerated
  particle exhibited a power-law like spectrum $N(\gamma) \propto
  \gamma^{-p}$ with a spectral index of roughly $p \simeq 2$.  

\begin{figure}[htb!]
   \plotone{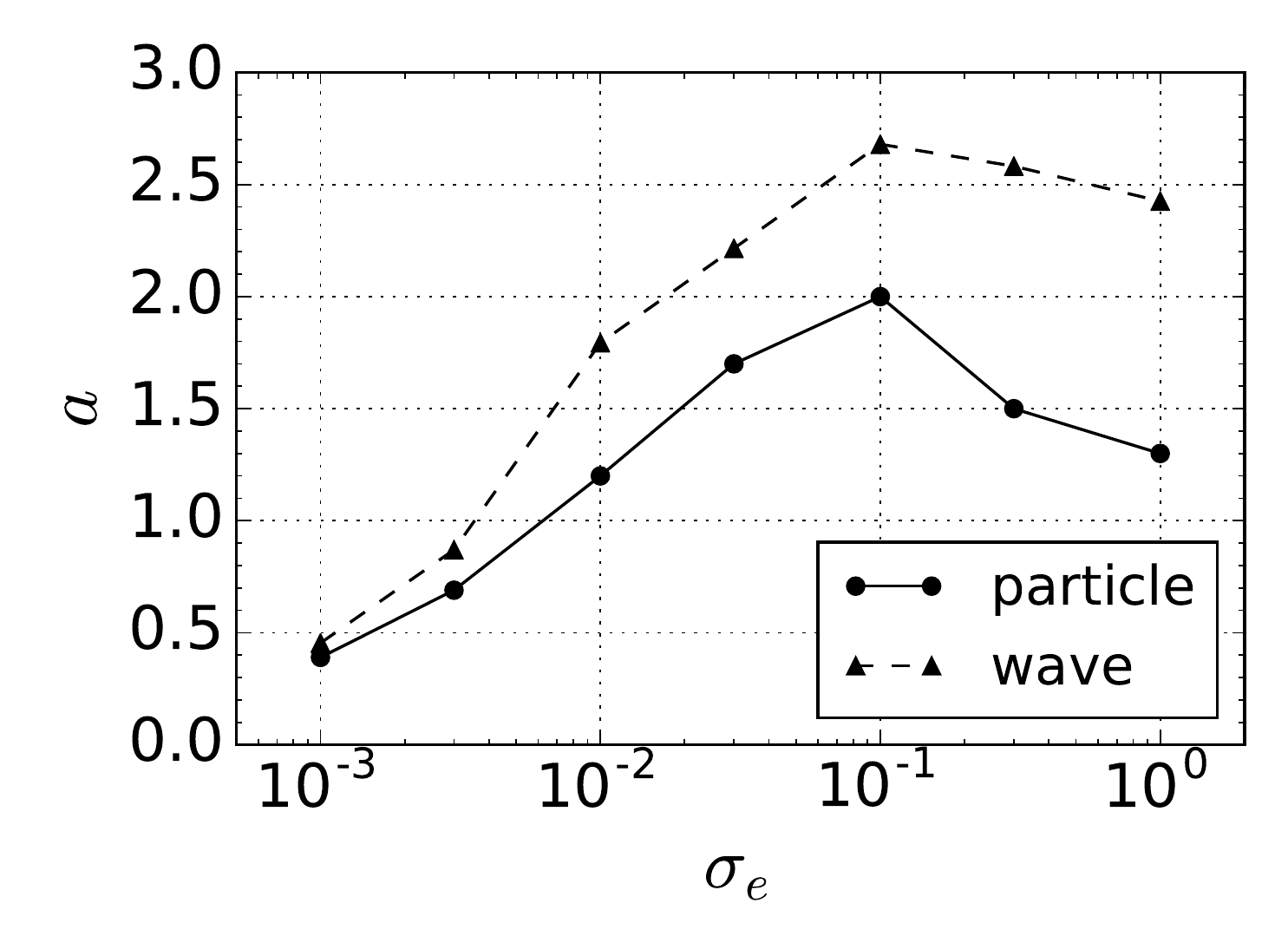}
   \caption{$\sigma_e$ dependence of the strength parameters estimated by
   the amplitudes of electron quiver motion (solid line) and precursor
   waves (dashed line). } 
   \label{strength}
  \end{figure}

  To discuss the relation with the results by \cite{Kuramitsu2008}, we
  estimated the strength parameter of the precursor waves from our
  simulation results. As the strength parameter is the amplitude
  of the particle quiver motion under the wave electromagnetic field, it may
  be estimated from the transverse particle velocity in the upstream
  region. As discussed in $\S$\ref{subsec:heating}, the particles in
  the upstream region are forced to oscillate in velocity by the strong
  electromagnetic field of the precursors. In the precursor region
  defined in $\S$\ref{subsec:highsig}, the oscillation amplitude of the
  particle velocity was determined by integrating the Fourier spectrum
  of the transverse velocity (i.e., first-order velocity moment)
  fluctuations over the wavenumber at $\omega_{pe} = 500$. The obtained
  strength parameter is shown in 
  Figure \ref{strength} by the solid line. Another way to estimate the strength 
  parameter is to use the precursor wave amplitude. Typical wavenumbers
  $kc/\omega_{pe} = 2-4$ of the observed precursor wave combined with the
  dispersion relation give the wave frequency as follows:
  \begin{equation}
   \frac{\omega}{\omega_{pe}} \sim 3.
  \end{equation}
  The strength parameter $a$ may then be estimated as
  \begin{equation}
   \label{eq:a_wave}
   a = \gamma_1 \sqrt{\sigma_e} \frac{\omega_{pe}}{\omega}\frac{\delta E}{B_1}
    \simeq  \gamma_1 \sqrt{\sigma_e} \frac{\omega_{pe}}{\omega}\frac{\delta
    B}{B_1} \sim \frac{\sqrt{2}}{3}\gamma_1\sqrt{\epsilon_{conv}}, 
  \end{equation}
  where $\epsilon_{conv} = \delta B^2/ 8\pi N_1\gamma_1 m_e c^2$ is the
  energy conversion rate from the upstream plasma kinetic energy into the
  precursor wave energy. 
  Figure \ref{strength} also shows
  the strength parameter estimated in this manner by the dashed line. It is
  readily apparent that the former method (based on the particle quiver velocity)
  gives a smaller value than the latter method (based on the wave
  amplitude). This is 
  natural because the latter estimate (Eq. \ref{eq:a_wave}) uses
  $\epsilon_{conv}$ as a proxy 
  of the amplitude of a monochromatic wave, whereas the actual spectra
  are rather broadband in wavenumber. Nevertheless, it is worth noting that
  both estimates give the same order of magnitude and essentially the
  same tendency. The strength parameter $a \gtrsim 1 $ implies that the
  precursor waves are strongly unstable against parametric
  instability. The formation of the observed density filaments is
  consistent with this understanding.
  
  We again emphasize that the WFA should not
  occur in a pair plasma. However, the observation of a strong
  precursor-plasma interaction in the upstream region indicates that the same
  precursors will generate large-amplitude Langmuir waves  if
  propagating in an ion-electron plasma. 

  The latter estimate predicts that the strength parameter of the
  precursor waves linearly scales with the shock Lorentz
  factor. Therefore, highly-relativistic shocks associated with, for
  instance, external shocks of GRBs may be an important site
  of the application of the WFA model.

 \section{Summary}\label{sec:summary}
 
 In this study, we investigated the efficiency of coherent
 electromagnetic precursor wave emission by the SMI at relativistic
 magnetized shocks in pair plasmas via 2D PIC simulations. The precursor
 wave emission 
 efficiency was measured as a function of the magnetization parameter
 $\sigma_e$, which well characterizes the relativistic shocks. Although
 the wave power in 2D simulations was systematically smaller than that
 in the corresponding 1D results, it was sufficiently strong to
 produce large-amplitude density filaments in the upstream medium. We
 think that the formation of density filaments is due to a filamentation
 instability excited by intense electromagnetic waves propagating in the
 upstream plasma. The precursor wave emission continues even well after
 the highly-disturbed upstream medium interacts with the shock, and the
 amplitude reaches a quasi-steady state level. At low $\sigma_e$,
 the power was roughly an order of magnitude 
 smaller than that in 1D simulations, which may be attributed to the
 presence of the WI 
 dominating the shock transition region. Nevertheless, we 
 found that large-amplitude precursor waves persist even in the
 Weibel-dominated regime. Therefore, we conclude that the emission of
 the coherent precursor waves is indeed intrinsic to relativistic
 shocks, even if the multidimensional effect is considered.

 Based on the simulation results, we discussed the applicability of the
 WFA model to astrophysical relativistic shocks. We concluded that the
 precursor wave power may be sufficiently strong for the WFA at
 highly-relativistic shocks in an ion-electron plasma. External shocks
 in the relativistic jets from GRBs may be important sites for the production
 of UHECRs via the WFA. However, the actual particle acceleration
 efficiency must be comprehensively examined by directly performing
 simulations for relativistic ion-electron shocks. This will be a
 subject for future research.
 
 \acknowledgments
 
 Numerical computations were carried out on Cray XC30 at Center for
 Computational Astrophysics, National Astronomical Observatory of
 Japan.

 \appendix

 \section{Numerical Convergence}\label{sec:conv}

 We also performed 1D PIC simulations for numerical convergence
 study. The simulation setup was identical to that 
 described in $\S$\ref{sec:setup}, except for the number of grids and
 particles per cell. We used the same 2D simulation code with two grid
 points in the $y$ direction. For technical reasons, this is the
 minimum number of grid points in the transverse direction.

 First, we investigated the dependence on the number of particles per
 cell $N_1\Delta x$ for each species and found that the results did not
 change as long as $N_1\Delta x \gtrsim 64$. However, for a smaller
 number of particles, the measured amplitude  
 shows non-negligible decrease, probably due to enhanced numerical
 collisions. In general, the effect of numerical
 collision becomes smaller as the dimensionality increases
 \citep{Kato2013, May2014}. Therefore, $N_1 \Delta x = 64$ is
 sufficient for 2D simulations. 

 Next, we studied the dependence on the grid size $\Delta x$. For this
 study, we fixed the box size to  $N_x \Delta x/(c/\omega_{pe}) = 500$ where
 $N_x$ is the number of gird points. For each $\sigma_e$, we performed
 simulations for the following four different resolutions: 
 $\Delta x/(c/\omega_{pe}) = 1/10$, $1/20$, $1/40$ and $1/80$.  
 
\begin{figure*}[htb!]
  \plottwo{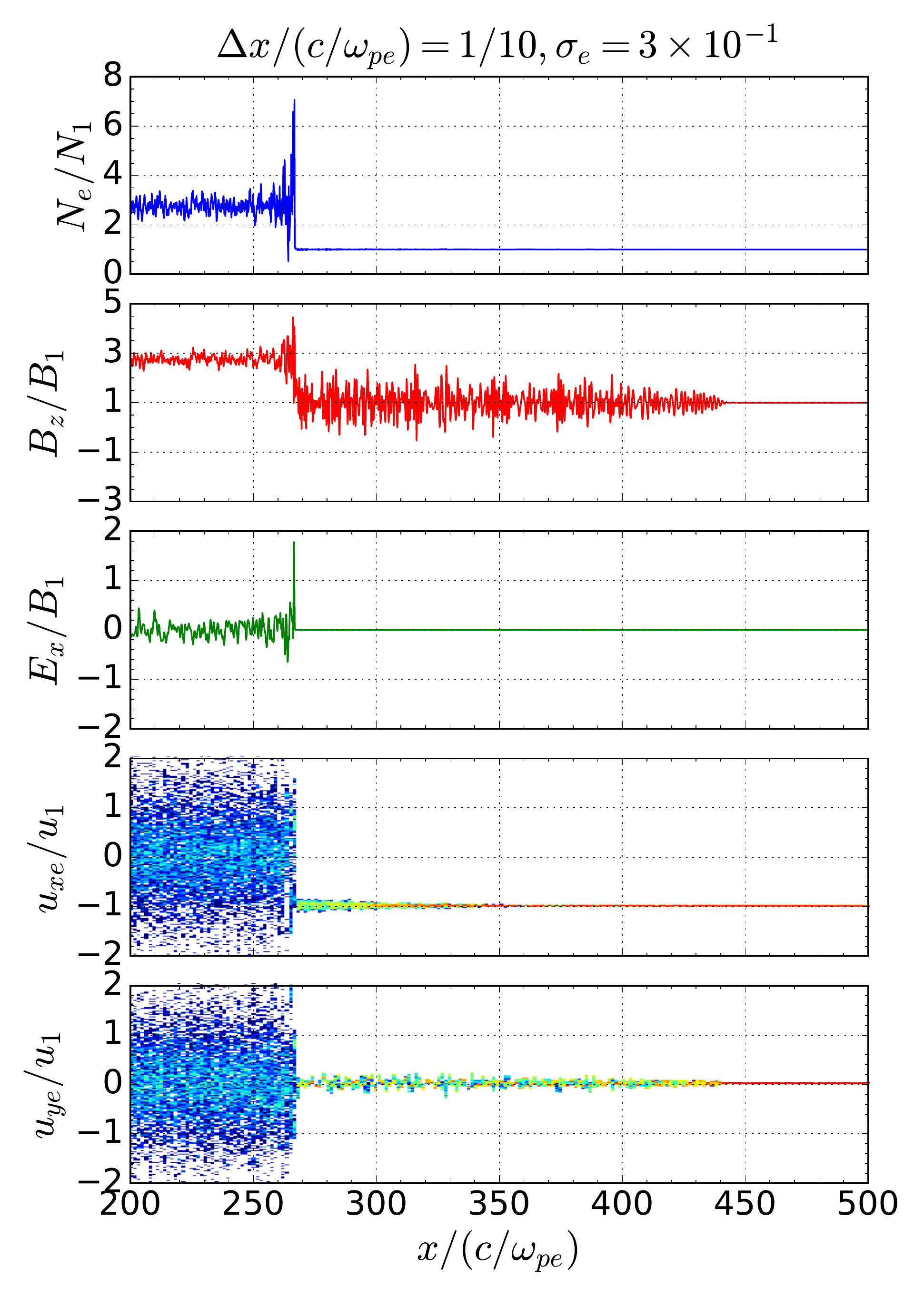}{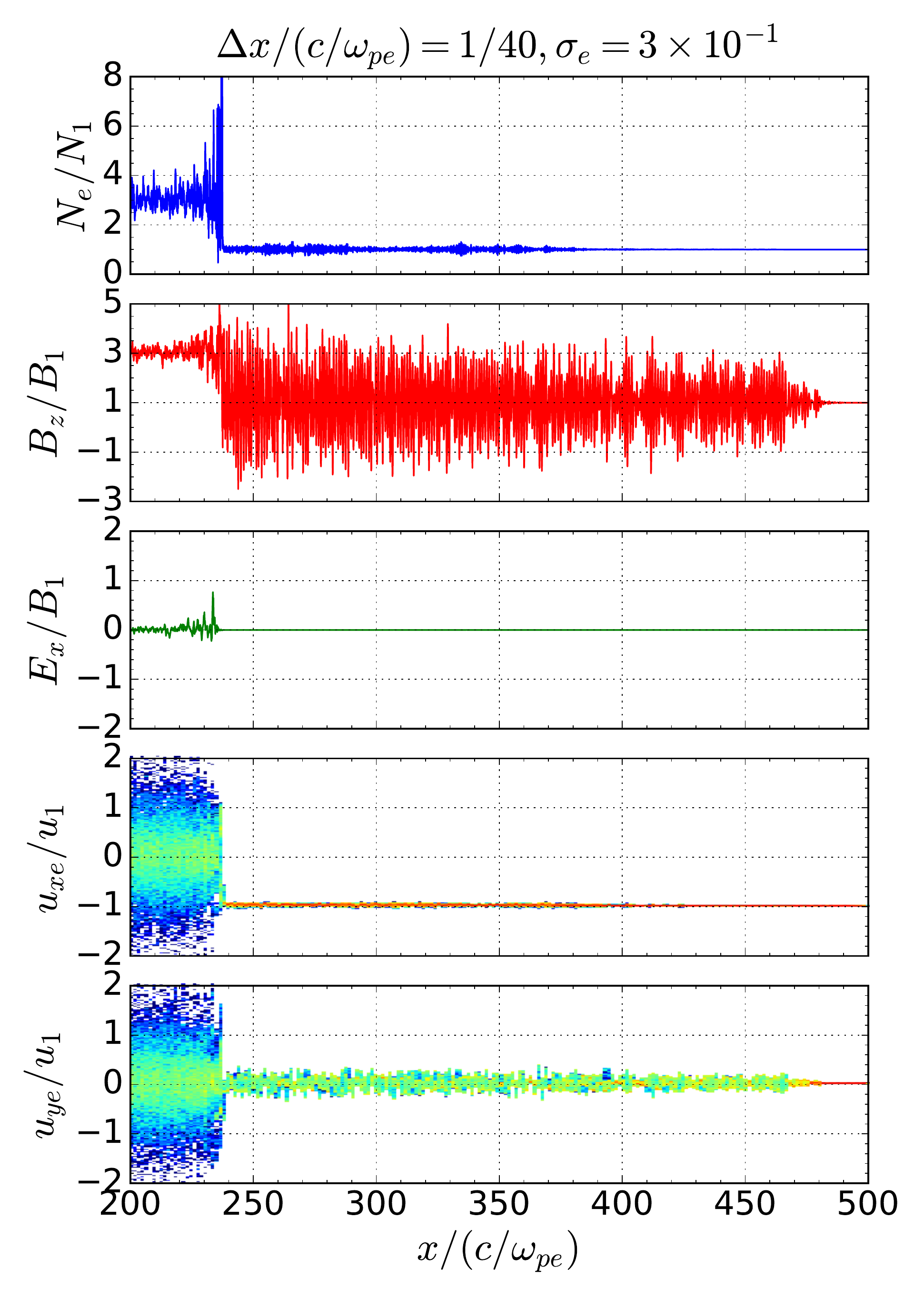}
  \caption{Comparison of 1D PIC simulations for $\sigma = 3 \times
  10^{-1}$ at two different resolutions: $\Delta x/(c/\omega_{pe})
  = 1/10$ (left) and $\Delta x/(c/\omega_{pe}) = 1/40$ (right). Top to
  bottom: electron number density $N_e$, $z$ component of 
  magnetic field $B_z$, $x$ component of electric field $E_x$, and
  phase-space plots in $x-u_x$ and $x-u_y$ at $\omega_{pe}t = 500$,
  respectively.}  
  \label{1dfield}
 \end{figure*}

 Figure \ref{1dfield} shows the overall shock structure at 
 $\omega_{pe}t = 500$ for $\sigma_e = 3 \times  10^{-1}$ for $\Delta x/
 (c/\omega_{pe})=1/10$ (left) and $\Delta x/(c/\omega_{pe})=1/40$
 (right). Shown from top to bottom are the electron number density $N_e$,
 the $z$ component of magnetic field $B_z$, the $x$ component of the
 electric field $E_x$, and the phase-space plots in $x-u_x$ and $x-u_y$,
 respectively. All quantities are normalized by the corresponding
 upstream values. It is clear that the positions of the shock front are
 different between the two cases: at $x/(c/\omega_{pe}) \sim 270$ for $\Delta
 x/(c/\omega_{pe})=1/10$ and at $x/(c/\omega_{pe}) \sim 240$ for $\Delta
 x/(c/\omega_{pe})=1/40$. We think that the difference in
 shock propagation velocity arises due to different precursor wave emission
 efficiencies, as shown in Figure \ref{1dfield}. The increased
 precursor wave emission indicates a larger fraction of the energy flux
 carried away from the shock, thereby modifying the Rankine-Hugoniot
 relation. In Appendix \ref{sec:cutoff}, we show actually that the
 precursor emission has substantial effect on the shock propagation
 speed in this parameter range. In contrast, the precursor
 region ahead of the shock is more extended in higher resolution runs
 than in lower resolution runs. This is in contrast to the
 shock propagation speed, and may be explained by analyzing the emitted
 wave spectrum.  

\begin{figure*}[htb!]
  \plottwo{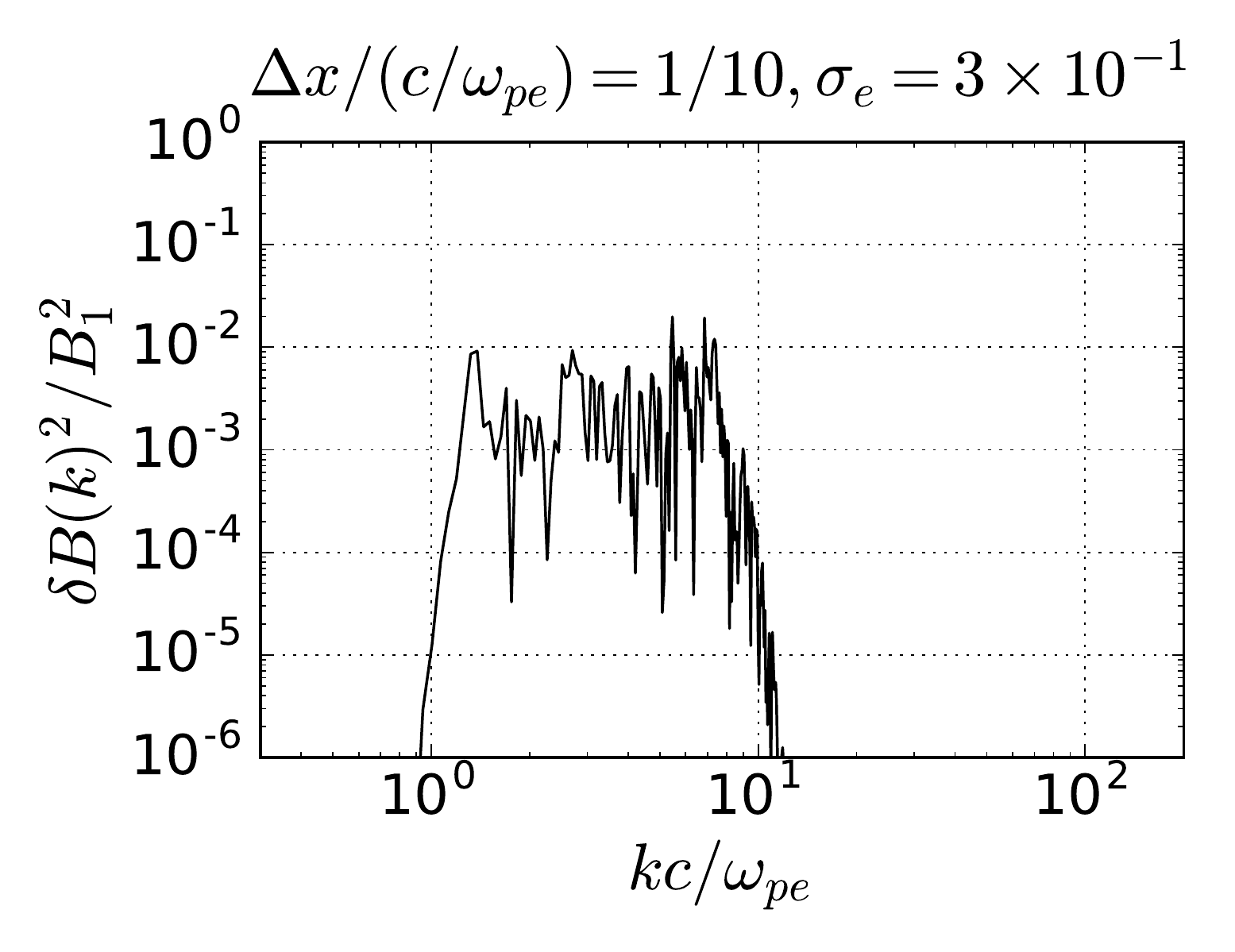}{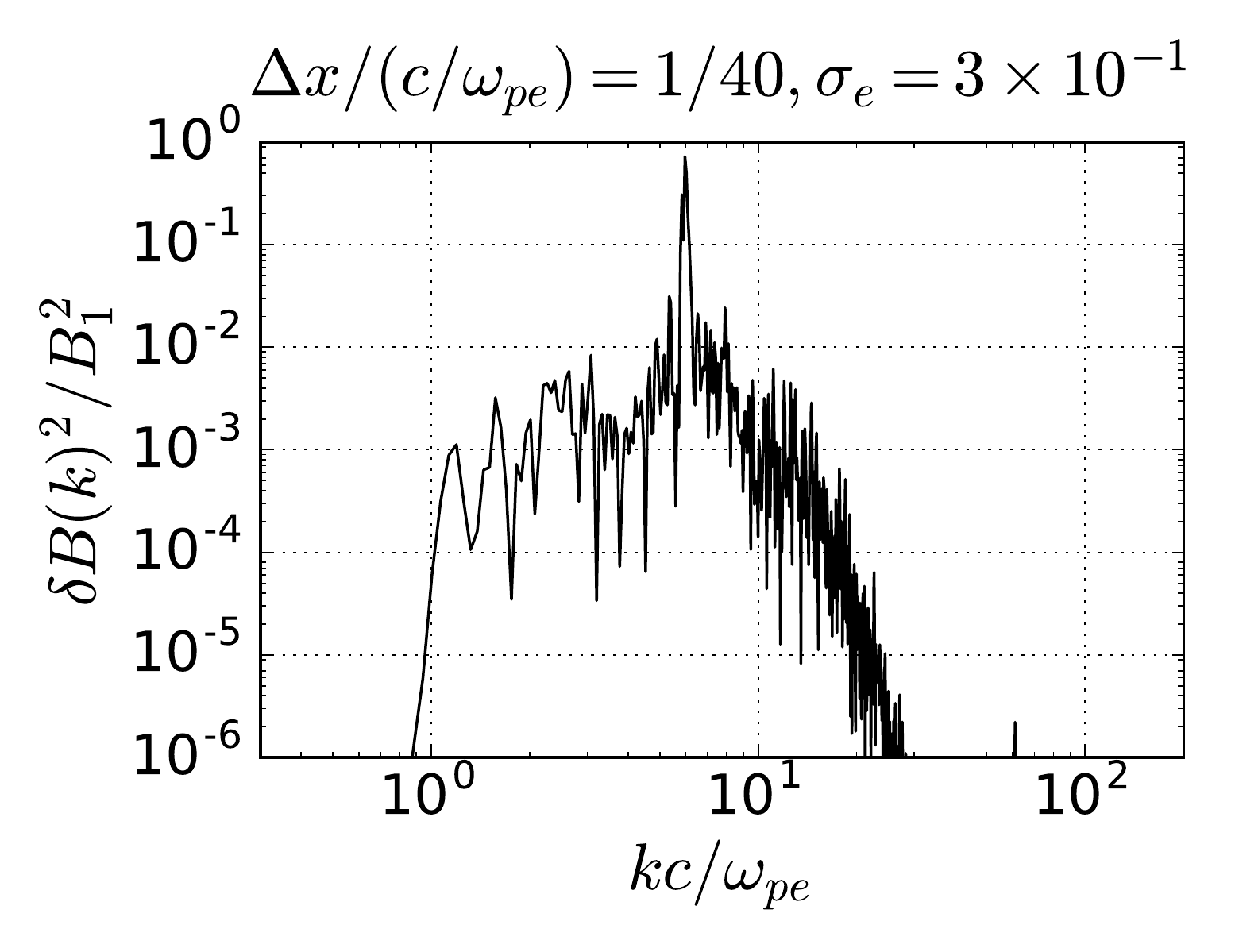}
  \caption{Wavenumber power spectra for precursor waves normalized by upstream
  ambient magnetic field at $\omega_{pe}t = 500$ for $\sigma_e = 3
  \times 10^{-1}$: $\Delta x/(c/\omega_{pe})=1/10$ (left) and $\Delta
  x /(c/\omega_{pe})=1/40$ (right).}
  \label{1dspectrum}
 \end{figure*}

 The power spectra of the precursor waves are shown in Figure
 \ref{1dspectrum}. The spectra were obtained for snapshots at 
 $\omega_{pe}t = 500 $ by taking the Fourier transform in the region
 $r_L < x-X_{sh} < r_L + 50c/\omega_{pe}$. The Nyquist wavenumbers are
 $k_N \sim 30 \omega_{pe}/c$ and $k_N \sim 120 \omega_{pe}/c$ for $\Delta x/
 (c/\omega_{pe})=1/10$ and $\Delta x/(c/\omega_{pe})=1/40$, respectively.  
 Both the maximum wavenumber and the peak power are greater for $\Delta
 x/(c/\omega_{pe})=1/40$ than for $\Delta x/(c/\omega_{pe})=1/10$. Because
 the group velocity of electromagnetic waves increases with the wavenumber,
 the precursor waves in the higher resolution run can propagate farther
 away from the shock front than at lower resolution. In addition,
 numerical damping (which is more significant at lower resolution)
 may also contribute to waves during long-distance propagation.

\begin{figure}[htb!]
  \plotone{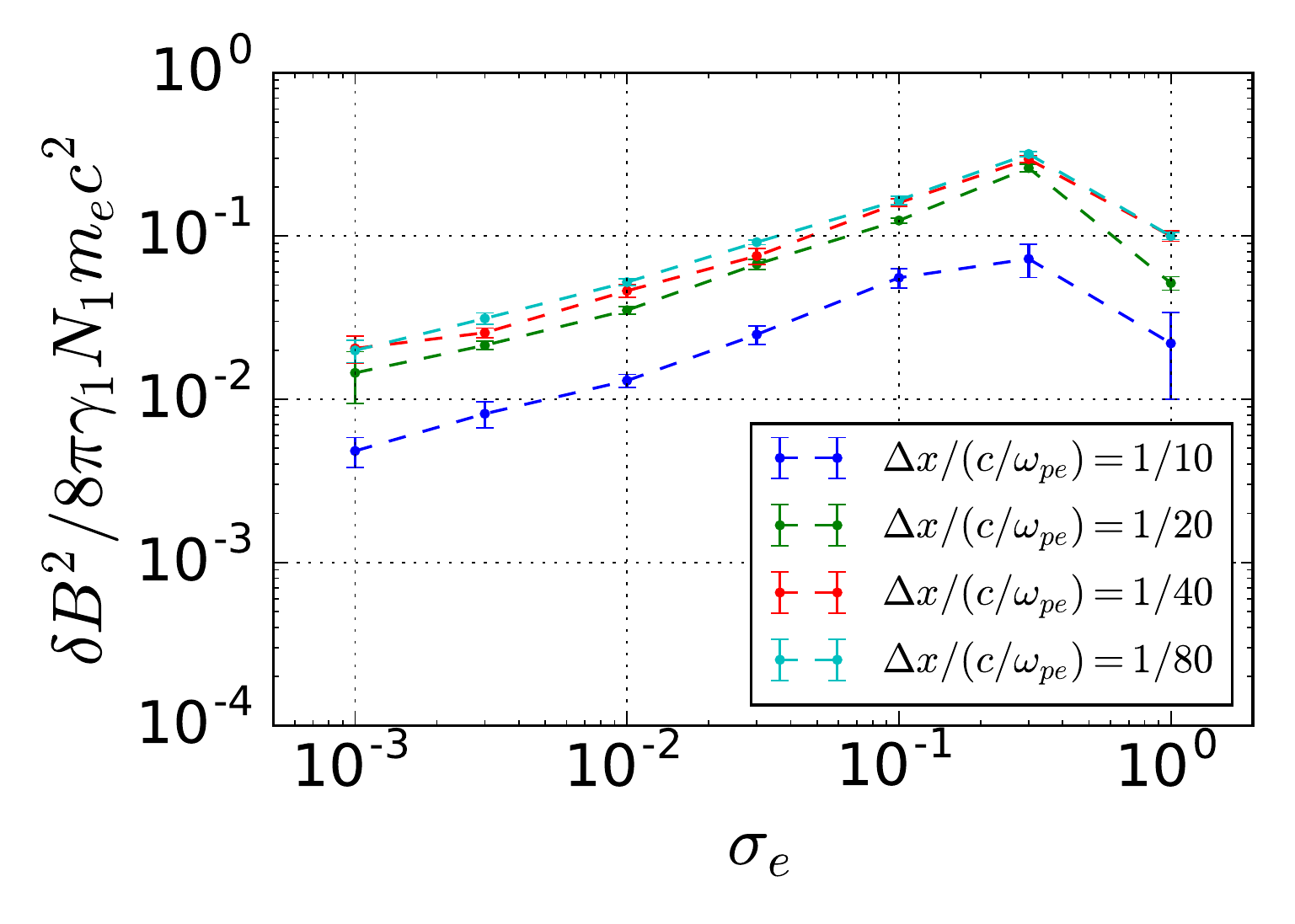}
  \caption{The $\sigma_e$ dependence of the precursor wave energy normalized
  by the upstream bulk kinetic energy. The results for $\Delta x/
  (c/\omega_{pe}) = 1/10, 1/20, 1/40$ and $1/80$ are shown by blue, green,
  red and cyan lines respectively.}
  \label{dx}
 \end{figure}

 Figure \ref{dx} shows the $\sigma_e$ dependence of the wave energy density
 normalized by the total upstream bulk kinetic energy for $\Delta x/
 (c/\omega_{pe}) = 1/10, 1/20, 1/40$ and $1/80$. The amplitude was calculated
 by integrating the power spectra over the whole wavenumber space. Error
 estimates were obtained by taking the standard deviation 
 during the time interval $ 500 \le \omega_{pe}t \le 520$. The wave
 amplitude systematically increases along with the resolution for all
 $\sigma_e$ and more or less converges for $\Delta x/ (c/\omega_{pe})
 \le 1/40$. This numerical convergence study confirms that 
 the precursor wave emission is indeed very sensitive to numerical
 resolution. Based on this result, we used  $\Delta x/ (c/\omega_{pe})
 = 1/40$ for the 2D simulations discussed in the main text.

 \section{Cutoff Wavenumber}\label{sec:cutoff}
 
  We here estimate a theoretical cutoff wavenumber, above which
  electromagnetic waves may escape from the shock upstream using the
  X-mode dispersion relation in a cold pair plasma. In the 
  plasma rest frame, this relation is given by
  \begin{equation}
   n^{\prime 2} = 1-\frac{2\omega_{pe}^2}{\omega^{\prime 2}-\sigma_e
    \omega_{pe}^2}, 
  \end{equation}
  where the prime indicates physical quantities measured in the plasma
  rest frame and $n^{\prime} \equiv c k^{\prime}/\omega^{\prime}$ is
  the refraction index. 
  By performing Lorentz transformation to the simulation frame, the
  dispersion relation upstream becomes
  \begin{equation}
   \label{eq:disp1}
    \frac{\omega^2}{\omega_{pe}^2} =
    \frac{2}{1-n^2}+\frac{\sigma_e}{\gamma_1^2(1+\beta_1n\cos \theta)^2},
  \end{equation}
  where $\theta$ is the angle between the wave propagation direction and
  the $x$ axis. As the precursor wave traveling upstream is considered,
  $\theta$ varies from $-\pi/2$ to $\pi/2$. Therefore, $1+\beta_1n\cos
  \theta$ is always greater than unity, and we can safely neglect the
  second term in Eq. \ref{eq:disp1} for $\sigma_e/\gamma_1^2 \ll 1$,
  which is satisfied in our simulations. This leads to the dispersion
  relation \ref{eq:disp1}, which is identical to that for a weakly magnetized
  plasma ($\sigma_e \ll 1$) in the plasma rest frame:
  \begin{equation}
   \label{eq:disp2}
    \omega^2 \simeq 2\omega_{pe}^2+k^2c^2.
  \end{equation}
  Now, the group velocity of the precursor wave can be expressed as
  \begin{equation}
   v_g \equiv \frac{d\omega}{dk} = \frac{kc^2}{\sqrt{2\omega_{pe}^2+k^2c^2}}.
  \end{equation}
  Equating the $x$ component of the group velocity $v_g \cos \theta$
  with the shock velocity $c\beta_{shock}$, we obtain the cutoff wavenumber
  \begin{equation}
   k_x = \beta_{shock}\gamma_{shock}\sqrt{k_y^2+\frac{2\omega_{pe}^2}{c^2}}.
  \end{equation}

  \begin{figure}[htb!]
   \plotone{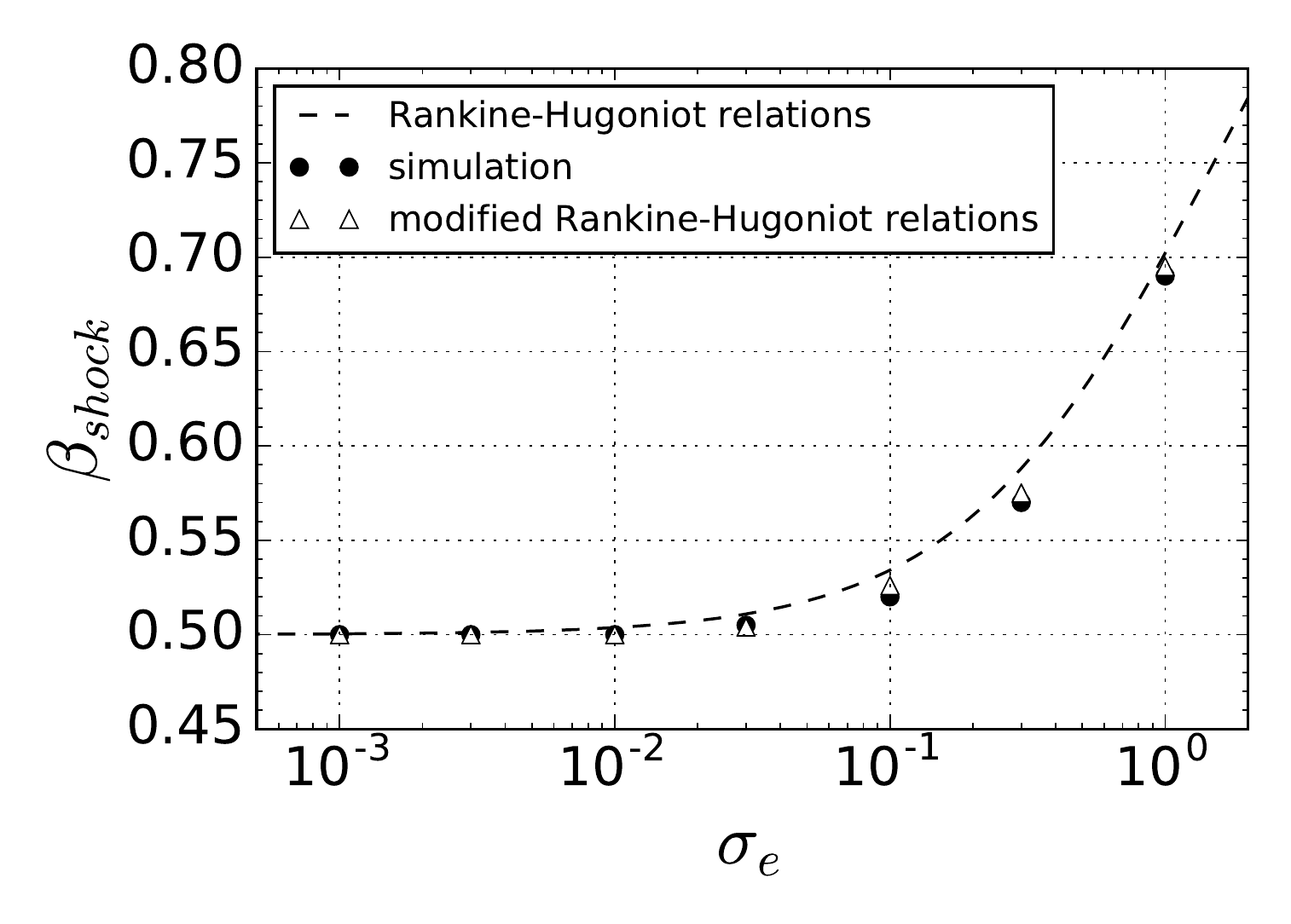}
   \caption{Comparison of the shock velocity measured in simulations
   with Rankine-Hugoniot relations. The dashed line shows the ideal MHD
   Rankine-Hugoniot prediction. The filled circles and open triangles
   show the measured shock velocity, and the shock velocity obtained by
   the modified Rankine-Hugoniot conditions, respectively.}  
   \label{rh}
  \end{figure}

  Determining the shock velocities is essential to calculate the
  theoretical cutoff wavenumber. The shock velocities were measured
  based on the time evolution of the electron number density averaged
  over the $y$ direction. The $\sigma_e$ dependence of the shock
  velocity is shown in Figure \ref{rh}. The filled circles are the
  simulation results, which clearly deviate, particularly at higher
  $\sigma_e$, from the dashed line, showing the ideal MHD
  Rankine-Hugoniot prediction calculated 
  with an adiabatic index of $\Gamma = 3/2$. This is because the
  electromagnetic precursor waves are not considered in the
  ideal MHD Rankine-Hugoniot condition. We examined the consistency
  between the simulations and theory considering the effect
  of precursor wave emission in the energy and momentum conservation
  laws. Namely, the shock velocity is calculated by solving the
  conservation laws using the energy and momentum fluxes 
  of the precursors in the simulations \citep[see][]{Gallant1992}. The open
  triangles in Figure \ref{rh} show the estimate based on the modified
  Rankine-Hugoniot conditions. This agrees quite well with the observed
  shock velocity, which we used to estimate the cutoff wavenumber.

 \section{Weibel Instability}\label{sec:weibel}

\begin{figure}[htb!]
  \plotone{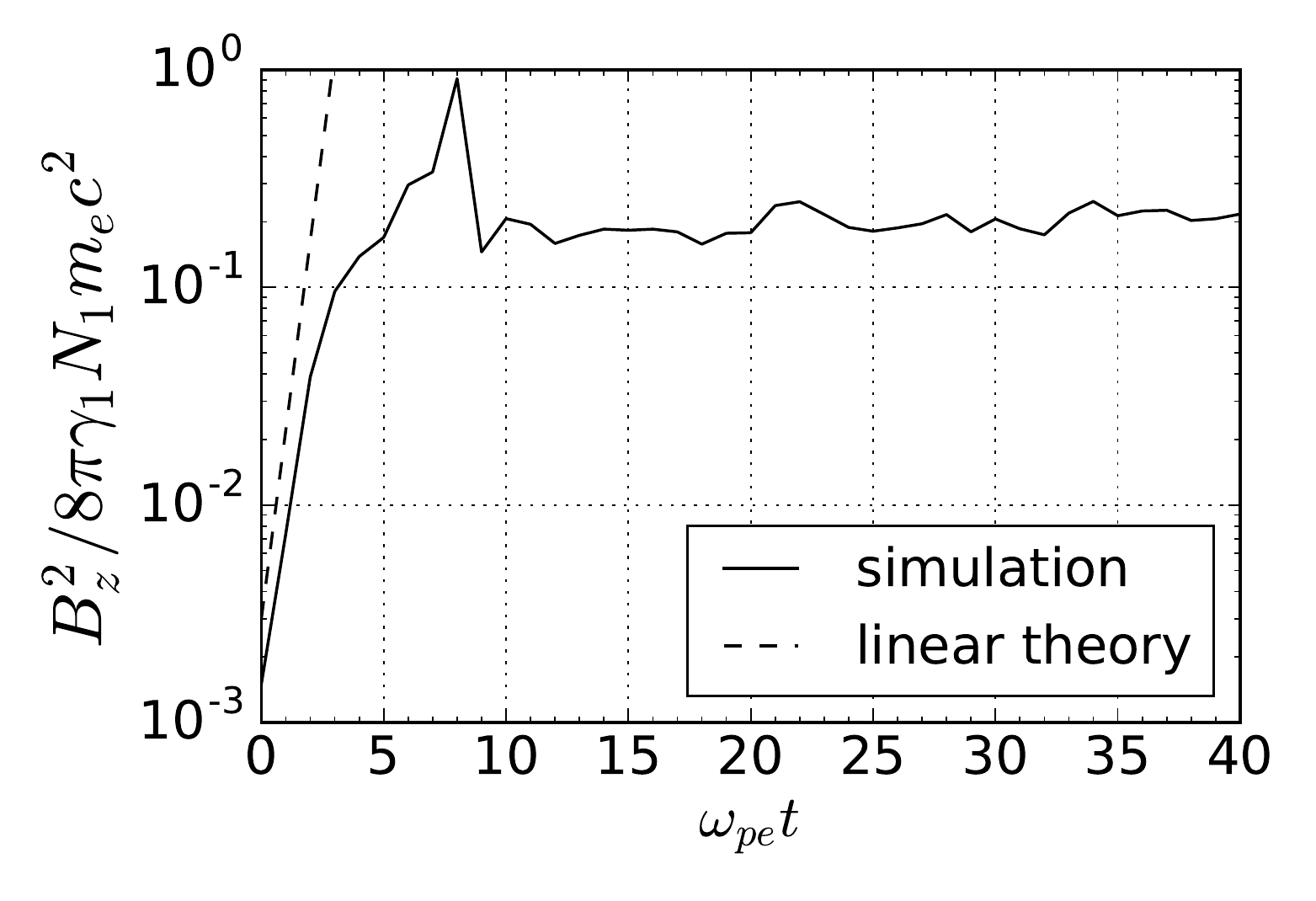}
  \caption{Time evolution of the magnetic field energy normalized by the
  upstream bulk kinetic energy. The solid line indicates the simulation
  results, whereas the dashed line indicates the prediction from the
  linear theory.}  
  \label{weibel}
 \end{figure}

 To confirm that the fluctuating magnetic fields near the shock front in
 the simulation with $\sigma_e = 3 \times 10^{-3}$ result from the WI,
 we here compare the time evolution of the magnetic field with the 
 linear growth rate of the WI. The maximum value of the fluctuating
 magnetic field energy averaged over the $y$ axis is determined for each
 snapshot (solid line in Figure \ref{weibel}). The
 fluctuating magnetic field energy is normalized by the upstream 
 bulk kinetic energy. For a low $\sigma_e$ shock wherein the WI
 substantially grows in amplitude, we may use a linear growth rate obtained for an
 unmagnetized plasma. \cite{Schaefer2008} discussed the linear theory of the
 WI for the monochromatic, waterbag, bi-Maxwellian and $\kappa$
 distribution functions. In all these distributions, the dispersion
 relation reduces to 
 \begin{equation}
  \label{eq:disp3}
  \omega^4 - (2\omega_{pe}^2+c^2k^2)\omega^2-\omega_{pe}^2c^2k^2 = 0
 \end{equation}
 for sufficiently strong anisotropy. Note that the relativistic effect
 was appropriately considered when calculating the linear growth
 rate. The maximum linear growth rate 
 $\Gamma_{max}$ can be estimated from the dispersion relation \ref{eq:disp3}
 \begin{equation}
  \Gamma_{max} \sim \omega_{pe},
 \end{equation}
 which is indicated in Figure \ref{weibel} by the dashed line. We see that the
 linear theory is consistent with the simulation result. The energy
 density of the generated magnetic field saturates at around $10-20\%$ 
 of the upstream bulk kinetic energy. This result is also consistent
 with those of the previous studies \citep{Kato2007, Chang2008,
 Sironi2011}. Therefore, we conclude that the filamentary magnetic field
 structure in the shock transition region results from the WI. 

\listofchanges

\end{document}